\documentclass[10pt,preprint]{emulateapj}

\def\thefootnote{\fnsymbol{ctr}}
\long\def\symbolfootnote[#1]#2{\begingroup%
\def\thefootnote{\fnsymbol{footnote}}\footnote[#1]{#2}\endgroup} 

\shorttitle{The Evolution of Massive Galaxies from $z\sim6$ to $z\sim1$}
\shortauthors{Lundgren et al.}

\begin{document}


\title{Tracing the Mass Growth and Star Formation Rate Evolution of Massive Galaxies from $z\sim6$ to $z\sim1$ in the Hubble Ultra-Deep Field}


\author{Britt F. Lundgren\altaffilmark{1,2}, Pieter van Dokkum\altaffilmark{3}, Marijn Franx \altaffilmark{4}, Ivo Labbe \altaffilmark{4,5}, Michele Trenti \altaffilmark{6,7}, Rychard Bouwens\altaffilmark{4}, Valentino Gonzalez\altaffilmark{8}, Garth Illingworth \altaffilmark{9}, Daniel Magee\altaffilmark{9}, Pascal Oesch \altaffilmark{3,10}, Massimo Stiavelli \altaffilmark{11}}





\altaffiltext{1}{Department of Astronomy, University of Wisconsin, Madison, WI 53706, USA}
\altaffiltext{2}{NSF Astronomy and Astrophysics Postdoctoral Fellow}
\altaffiltext{3}{Astronomy Department, Yale University, New Haven, CT 06511, USA}
\altaffiltext{4}{Leiden Observatory, Leiden University, NL-2300 RA Leiden, The Netherlands}
\altaffiltext{5}{Carnegie Observatories, Pasadena, CA 91101, USA}
\altaffiltext{6}{Institute of Astronomy, University of Cambridge, Madingley
Road, Cambridge, CB3 0HA, United Kingdom}
\altaffiltext{7}{Kavli Institute for Cosmology, University of Cambridge,
Madingley Road, Cambridge, CB3 0HA, United Kingdom}
\altaffiltext{8}{Department of Physics and Astronomy, University of California, Riverside, CA 92521, USA}
\altaffiltext{9}{UCO/Lick Observatory, University of California, Santa Cruz, CA 95064, USA}
\altaffiltext{10}{Hubble Fellow}
\altaffiltext{11}{Space Telescope Science Institute, 3700 San Martin Drive
Baltimore MD 21218 USA}


\begin{abstract}
We present an analysis of an $H_{160}$-selected photometric catalog of galaxies in the Hubble Ultra-Deep Field, using imaging from the WFC3/IR camera on the HST in combination with archival UV, optical, and NIR imaging.  Using these data we measure the spectral energy distributions of $\sim1500$ galaxies to a limiting $H_{160}$ magnitude of 27.8, from which we fit photometric redshifts and stellar population estimates for all galaxies with well-determined {\it Spitzer} IRAC fluxes, allowing for the determination of the cumulative mass function within the range $1<z<6$.  By selecting samples of galaxies at a constant cumulative number density, we are able to explore the co-evolution of stellar masses and star formation rates for progenitor galaxies and their descendants from $z\sim6$.  We find a steady increase in the star formation rates of galaxies at constant number density from $z\sim6$ to $z\sim3$, accompanied by gradually declining specific star formation rates during this same period.  The peak epoch of star formation is also found to shift to later times for galaxies with increasing number densities, in agreement with the expectations from cosmic downsizing. The observed star formation rates can fully account for the mass growth to $z\sim2$ amongst galaxies with cumulative number densities greater than 10$^{-3.5}$ Mpc$^{-3}$.  For galaxies with a lower constant number density (higher mean mass), we find the observed stellar masses are $\sim3$ times greater than that which may be accounted for by the observed star formation alone at late times, implying that growth from mergers plays an important role at $z<2$.  We additionally observe a decreasing specific star formation rate, equivalent to approximately one order of magnitude, from $z\sim6$ to $z\sim2$ amongst galaxies with number densities less than 10$^{-3.5}$ Mpc$^{-3}$ along with significant evidence that at any redshift the specific star formation rate is higher for galaxies at higher number density.  The combination of these findings can qualitatively explain the previous findings of a sSFR plateau at high redshift. Tracing the evolution of the fraction of quiescent galaxies for samples matched in cumulative number density over this redshift range, we find no unambiguous examples of quiescent galaxies at $z>4$. 

\end{abstract}


\keywords{galaxy evolution: general}



\section{Introduction}


Owing to the successful addition of the WFC3/IR camera on the Hubble Space Telescope, new deepest-ever views of the Universe are providing fresh insight into the physical properties of high-redshift galaxies.  Infrared images of the Hubble Ultra-Deep Field \citep[HUDF;][]{Beckwith06} have already enabled detections of galaxies with photometric redshifts in excess of $z\sim8$ \citep[e.g.,][]{Bouwens2010b, Oesch12}, from which the UV-luminosity functions of extremely high-redshift galaxies have been measured for the first time.  While these extraordinary measurements of infrared drop-out selected galaxies provide much needed information about the most actively star-forming objects at high-redshift, a more comprehensive multi-wavelength analysis is required to study the evolution of typical galaxies from early epochs to the present.

Until recently, observational studies of galaxy evolution have struggled to find unbiased methods of comparing galaxies over wide ranges in redshift.  Evolution in the star-formation histories, mass, and luminosities of populations of galaxies over a broad range in redshift mean that selection of galaxies by luminosity, color, or mass produce significant bias in evolution studies \citep[e.g.,][]{PvD06}.  Recent works have demonstrated the effectiveness of selecting similar galaxy populations with constant number density, which enabled the study of similar samples of galaxies from $z\sim2$ to present \citep{CW09, PvD2010}.

\begin{table*}
\centering
\begin{minipage}{0.75\textwidth}
\begin{center}
\caption{Summary of Included HUDF Observations}
\begin{tabular}{@{}rrrrrrr@{}}
\hline
Camera & Filter  & $\lambda_{c}$ & FWHM & Zeropoint & Limiting   \\
& &  ($\AA$) & (\arcsec) & (mag) & Magnitude$^{\footnote{Limiting magnitudes for the ACS and WFC3/IR images are determined at 5$\sigma$ for a 0.\arcsec35 diameter aperture; VIMOS depths are measured at 5$\sigma$ within a 1.\arcsec2 diameter aperture; IRAC depths are quoted for a 1$\sigma$ depth in a 2.\arcsec5 aperture.}}$ \\
\hline
\hline
VIMOS & $U_{38}$ & 3656 & 0.8 & 26.150 & 28.6\\
 \hline
 ACS & F435W & 4327 & 0.08 & 35.635 & 29.7 \\
 & F606W & 5957 & 0.08 & 36.460 & 30.1 \\
 & F775W &  7705 & 0.09 & 35.636 & 29.9 \\
 & F850LP & 9072 & 0.09 & 34.847 & 29.4 \\
 \hline
 WFC3/IR & F105W & 10580 & 0.14 & 33.758 & 29.6 \\ 
 & F125W & 12490 &  0.15 & 33.742 & 29.9\\
 & F160W & 15440 & 0.16 & 33.455 & 29.9\\
 \hline
 ISAAC/PANIC & $K_{s}$ & 21681 & 0.36 & 26.150 & 27.4\\
 \hline
 IRAC & ch1 & 35634 & 1.7 & 22.416 & 27.7 \\
 & ch2 & 45110 & 1.7 & 22.195 & 27.2 \\
\hline
\end{tabular}
\end{center}
\end{minipage}
\end{table*}

By stacking rest-frame R-band images of galaxies selected at a constant number density of $n\sim2\times10^{-4}$ Mpc$^{-3}$ \citet{PvD2010} explored the stellar mass and radius evolution of massive galaxies from $z=2$ to $z=0$.  The results indicated that massive galaxies form in an inside-out fashion,  in which the outer regions assemble at later times around a dense, compact core.  In-situ star formation could only account for $\sim20$\% of the observed mass growth, implying that mergers dominate other contributions to the mass growth of galaxies at this number density.  

\citet{Papovich11}, extended the analysis of \citet{PvD2010} to higher redshift to trace the evolution of galaxies with a constant cumulative number density of 2$\times10^{-4}$ Mpc$^{-3}$ , as implied by the combined luminosity functions of \citet{RS09}, \citet{Bouwens07}, and \citet{Bouwens2010a}.  By exploiting the empirical relation between mass and luminosity, \citet{Papovich11} reconstructed the cosmologically averaged stellar mass growth for galaxies with this fixed number density over the range $3<z<8$.  Comparing this stellar mass growth to the star formation rates implied by the luminosity functions, \citet{Papovich11} determined that the growth could be entirely accounted for by star formation.  Moreover, they reported a steady increase in star formation rates for galaxies with this fixed number density from $z\sim8$ to $z\sim3$.  This result, which is predicted by hydrodynamic and semi-analytic simulations, implies that cosmically-averaged gas accretion rates at $z>4$ are at least as high as the concurrent star formation rates, making gas accretion a dominant source of mass growth in this high-redshift regime.

Testing the results of \citet{Papovich11}, which were derived with a rest-frame UV-selected luminosity function, with direct measurements from a stellar mass-limited sample of individual galaxies is a necessary next step for better understanding the evolution of massive galaxies at high-z, which is now possible using a combination of new and archival multi-wavelength observations available in the HUDF.  A mass-selected sample enables a more complete measure of the mean stellar mass build-up in samples of galaxies with constant number density, since it achieves higher completeness for galaxies with lower star formation rates.

We present a multi-wavelength photometric catalog of $\sim1500$ infrared-selected  galaxies in the HUDF, which incorporates deep UV, optical, and infrared imaging.   Using precisely determined spectral energy distributions (SEDs) covering both the rest-frame UV and optical, we estimate the star formation histories and stellar masses of populations with constant cumulative number density from $z\sim6$ to $z\sim1$.   We additionally supplement the high-redshift measurements in the HUDF with data from the larger overlapping FIREWORKS catalog \citep{Wuyts08} in order to extend our study of the evolution of high-mass high-redshift galaxies to $z\sim0.5$.  

The observations contributing to the HUDF galaxy catalog, which we provide as a supplement to this work, are detailed in Section 2.  The image reduction, source extraction, and photometry are described in Section 3 along with details of the photometric redshift estimates and stellar population modeling.  Section 4 describes the method we apply for producing samples of galaxies with constant cumulative number density, and in Section 5 we present the average evolution of these galaxy samples as measured in stellar mass, star formation rate, and specific star formation rate.  Throughout this paper, we assume a flat $\Lambda$--dominated CDM cosmology with $\Omega_m=0.27$, $H_0=73$ km s$^{-1} $Mpc$^{-1}$, and $\sigma_8=0.8$ unless otherwise stated.  All magnitudes are provided in the AB system \citep{OG83}.

\section{Observations}

The Hubble Ultra-Deep Field \citep[HUDF;][]{Beckwith06}, is an 11 arcmin$^{2}$ region of the sky within the GOODS-South field, centered at 3$^{h}$32$^{m}$38$^{s}$.5 -27$^{d}$47'0$^{''}$.0.   This region of the sky was selected for its general absence of bright stars with the purpose of providing the deepest ever view of the Universe.  The deep imaging of the HUDF, initially provided by the Advanced Camera for Surveys (ACS) on the {\it Hubble Space Telescope (HST)} and later supplemented with multi-wavelength observations, has produced an exquisite laboratory for the study of galaxy evolution from high redshift.  In the following section we describe both the new and archival data used to produce the photometric galaxy catalog presented in this work.  A summary of these observations is provided in Table 1.

\subsection{The WFC3/IR $Y_{105}$, $J_{125}$, \& $H_{160}$ Data}

The galaxies we examine have been selected from new $H_{160}$-band imaging from the ultra-deep near-IR WFC3/IR observations in the main HUDF field, part of the 192-orbit HUDF09 program (PI Illingworth: G0 11563).  The HUDF09 program includes WFC3/IR images in three filter bands: $Y_{105}$, $J_{125}$, and $H_{160}$.  The resulting images reach limiting magnitudes of $\sim29$ mag (5$\sigma$, 0$\arcsec$.35 diameter apertures; see \citet{Bouwens2010a})  with a native pixel scale of 0.06\arcsec pixel$^{-1}$ and a PSF FWHM of $\sim0.16$\arcsec.  While deep imaging has also been obtained as part of the HUDF09 program for the flanking deep fields (HUDF09-1,2) within the larger HUDF05 (PI: Stiavelli), for the purpose of this analysis we incorporate only data from the primary HUDF09 field, which totals 4.7 arcmin$^{2}$.

\subsection{The ACS $B_{435}$ $V_{606}$ $i_{775}$ $z_{850}$ Data}

Imaging of the HUDF obtained with the ACS instrument on the {HST} include observations in four optical filter bands: F435W, F606W, F775W, and F850LP.  The observations were obtained in a total of 400 orbits between September 2003 and January 2004 \citep{Beckwith06}.  The co-added images, now publicly available, have been matched to the pixel scale of the new WFC3/IR data.  The respective limiting magnitudes of the $B_{435}$, $V_{606}$, $i_{775}$, and $z_{850}$ images are: 29.7, 30.1, 29.9, 29.4, as determined at 5$\sigma$ within a 0.\arcsec35 diameter aperture.

\subsection{VIMOS $U$-band Data}

We include deep $U$-band imaging taken with the VIMOS instrument on the VLT's Melipal Unit Telescope at Cerro Paranal Observatory, in Chile. These data were obtained by the ESO large program 168.A-0485 (P.I. C. Cesarsky), as part of the Great Observatories Origins Deep Survey (GOODS).   The individual frames contributing to the co-added image were obtained in service-mode between August 2004 and 2006. The combined $U$-band image reaches a 5$\sigma$ limiting magnitude of 28.6, measured in a 1.\arcsec 2 diameter aperture with an average point-spread function (PSF) FWHM of $\sim0\arcsec.8$ \citep{Nonino09}.

\subsection{The VLT/ISAAC $K_{s}$-band Data}

The $K_{s}$ band photometry in this work has been extracted from a combination of images taken with the Infrared Spectrometer and Array Camera (ISAAC) mounted on the Antu Unit Telescope of the Very Large Telescope (VLT) and the PANIC near-infrared camera on the 6.5m Magellan telescopes.  The VLT/ISAAC exposures were taken in two programs, one consisting of 8 hours on the GOODS Chandra Deep Field South, and another 15 hour exposure in the UDF.  The Magellan PANIC data includes 15 hours of exposure.   These observations have been combined and matched to the native pixel scale of the WFC3/IR images.  The final combined image reaches a limiting magnitude of 27.4, with an average PSF FWHM of $\sim0\arcsec.36$ (Labb{\'e} et al., in preparation).

\subsection{The IRAC $3.6,4.5 \mu m$ Data}

We include the super-deep observations from the GOODS-S Field, which were obtained using the Infrared Array Camera (IRAC; Fazio et al. 2004) on the $Spitzer$ $Space$ $Telescope$ were taken as part of the $Spitzer$ Legacy Program.  These data were taken with two pointings, each with an exposure time of approximately 23.3 hours. The $\sim40$ arcmin$^{2}$ of overlapping area with the HUDF received twice the exposure.  We include imaging from channels 1 and 2, which are centered on wavelengths of $3.6$ and $4.5 \mu m$, respectively, each with a 1$\sigma$ limiting depth of $\sim27$.

\section{Photometric Catalog Construction}

In the following section we describe the methods of source detection and photometry used to produce the multi-wavelength photometric catalog in the HUDF.  We also detail the estimation of photometric redshifts and the stellar population synthesis modeling using the determined SEDs.

\begin{figure*}[!ht]
\epsscale{0.75}
\begin{center}
\plotone{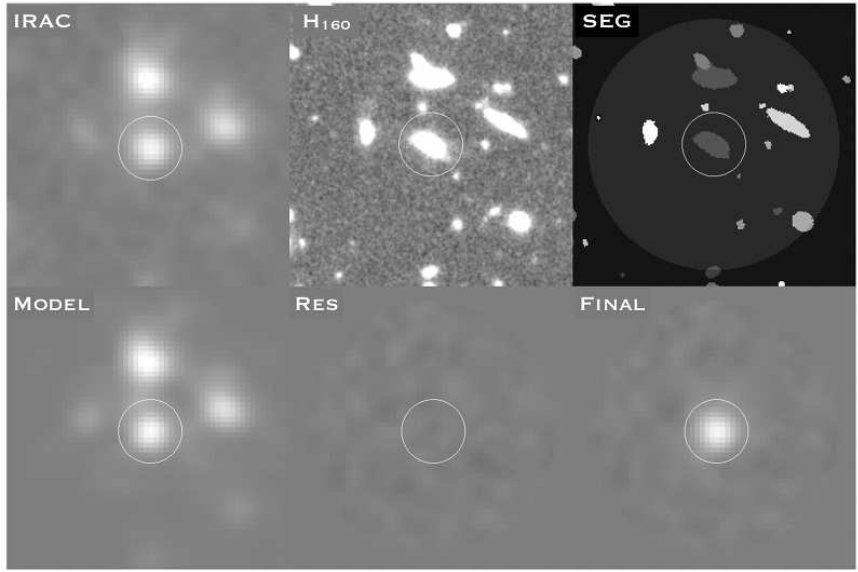}
\caption{A series of images depicting the process of modeling and deblending IRAC fluxes for objects identified in the $H_{160}$ detection image.  The original IRAC 3.6$\mu m$ cutout image for catalog object ID 764 is shown top left.  The top center panel gives the matching $H_{160}$ detection image, and the segmentation map of the region from \textsc{SExtractor} is shown at top right.  The bottom panel shows, from left to right, the modeled IRAC flux for all objects in the region, the residual image with all modeled fluxes removed, and the flux for the central object alone. \label{fig:subphot}}
\end{center}
\end{figure*}

\subsection{PSF Matching of Space-based Optical and IR Images}

The point spread function (PSF) varies significantly between the images included in this analysis,  such that the PSF FWHM generally increases with an increasing central wavelength of the broad-band imaging.  If left uncorrected, this variance would produce a wavelength-dependent bias in our aperture photometry.  

To alleviate this problem in the $HST$ data, we degrade each image to match the PSF of the image with the broadest PSF, which in this case is the $J_{125}$ image.  We begin by determining the PSF of the ACS and WFC3/IR images using a stack of seven non-saturated stars found in the field.  Any detected neighboring objects are masked to reduce background fluctuations in the stacked image.  
We produce kernels for the stellar composite in each band using the \textsc{lucy} algorithm in \textsc{IRAF}, and each image is convolved to match the $J_{125}$ PSF using the \textsc{IRAF} \textsc{fconvolve} routine.  The FWHMs of the PSFs in the resulting images match to within 2\%, as does the fraction of the total flux enclosed in a 10 pixel radius aperture.  

\subsection{PSF Matching for Other Bands}

Images with PSFs that are significantly broader than that of the $J_{125}$ detection image require more complicated PSF matching techniques.  For the VIMOS, K$_{s}$-band, and IRAC images, we therefore apply a different method, which allows for precise aperture photometry in cases where broad-winged PSFs can produce considerable contamination in the flux measurements of neighboring sources.  We employ the same source-fitting algorithm described in detail in \citet{Labbe06}, \citet{Wuyts08}, and \citet{Whitaker11}.  This method produces a model PSF for the image with the broader native PSF, which is then used to estimate the flux distribution of each source identified in the detection image segmentation map output by SExtractor, as described in the following section.  For each individual object, the flux from neighboring sources is modeled and subtracted, allowing for a reliable aperture flux measurement of individual objects, even in reasonably crowded areas of the field.  An example illustrating this PSF-matching technique, which we use to directly compare VIMOS, $K_{s}$-band, and IRAC images with the $H_{160}$ detection image, is shown in Figure~\ref{fig:subphot}.

In total, we measure K-band fluxes with a SNR $>2$ for 728 objects and 3.6$\mu$m fluxes with a SNR $>2$ for 326 objects.  For a small number of cases, particularly for objects in close proximity to very bright IRAC sources, blending problems are so severe that this procedure is unable to produce reasonable estimates of the flux in faint nearby objects.  By visual inspection, 34 such cases were identified in the photometric catalog.  Where present, this deblending failure generally produces an over-estimate of the flux in the $K_{s}$-band and IRAC bands.  Since the space-based optical and near-infrared photometry otherwise contains negligible contamination for most of these objects, they are retained in the catalog, but their fluxes in the $K_{s}$ and IRAC bands have been set to -99.  While these objects are retained in the catalog accompanying this work, we do not include them in the scientific analysis described in Section 5, as the $K_{s}$ and IRAC flux measurements are critical to the accuracy of stellar mass estimates.  The objects which we have removed from our analysis because of catastrophic deblending problems in the infrared represent just 2\% of the total dataset and are drawn from a random distribution in redshift.  Thus, we assume any correction to our analysis due to their removal is negligible.

\subsection{Source Extraction and Photometry}

We use the \textsc{SExtractor} code \citep{Bertin96} in dual-image mode to extract sources detected in the WFC3/IR $H_{160}$ image.  In order to minimize false detections while maximizing the efficiency of faint source detection we require a  minimum detection area of 5 pixels with a threshold of 5$\sigma$ significance above the background for both detection and analysis.  We apply 64 deblending sub-thresholds with a minimum contrast parameter of 0.002.  The background is estimated in \emph{GLOBAL} mode, using a mesh size of 100 pixels and a filter size of 3 pixels.

The initial catalog output from \textsc{SExtractor} contains 1645 objects, which is reduced to 1553 after removing objects with coverage of $<30$\% of the maximum in the $H_{160}$ weight map.  This cut effectively removes spurious detections near the edge of the WFC3/IR imaging.  An additional 8 objects are removed after being identified as stars or artifacts.

In order to maximize the average signal-to-noise, circular apertures with a 10 pixel radius, corresponding to 0$\arcsec$.6 at the pixel scale of the WFC3/IR images are used to measure color fluxes for all objects.  Total flux measurements are estimated in the $H_{160}$ detection image, using the \textsc{SExtractor} \emph{AUTO} photometry and the following criteria:
\begin{equation}
r_{aper}= \left \{ \begin{array}{rl}
r_{Kron},   &\mbox{$r_{Kron}>r_{color}$} \\
r_{color}, &\mbox{$r_{Kron}\leq r_{color}$} \label{eq:kron}
\end{array} \right.
\end{equation}
\begin{equation}
H_{total} = H_{AUTO} \times \frac{1}{F_{r<r_{aper}}} \label{eq:auto}
\end{equation}
where $r_{color}$=0\arcsec.6,  $r_{Kron}$ is the radius of the \textsc{SExtractor} circularized Kron aperture in units of arcseconds, and $F_{r<r_{aper}}$ is the aperture correction used to recover flux falling outside the total aperture radius, r$_{aper}$.  This correction is determined by the relative flux in the $H_{160}$ growth curve at the position of the circularized Kron radius of the AUTO aperture.  Measurements of the total flux in each of the additional HST imaging bands are calculated by scaling the 10 pixel aperture fluxes in the following manner:

\begin{equation}
F_{total} = F_{aper}\times \frac{H_{total}}{H_{aper}}\label{eq:apflux}
\end{equation}

We derive errors for the space-based photometry by measuring the 1$\sigma$ variance in a sampling of 2000 randomly placed circular apertures with a 10 pixel radius in each broad-band image.  Before placing the apertures, we subtract all detected objects, using the \textsc{SExtractor} segmentation map as a mask.  From the average error returned by this method, we determine the total error on the measurement of $F_{total}$ for each object, scaling up the estimated 1$\sigma$ color aperture error by the same factor of $\frac{H_{total}}{H_{aper}}$, used to calculate $F_{total}$ in Equation~\ref{eq:apflux}.   Errors on the photometry in the $U$, $K_{s}$, and \textsc{IRAC} bands are determined by a similar sampling of the residual source-subtracted model images (e.g., bottom center panel in Figure~\ref{fig:subphot}).   

For completeness, we include the catalog in the electronic version of the paper (see Appendix A).

\begin{figure}[t!]
\centering
\begin{center}
\epsscale{1.}
\plotone{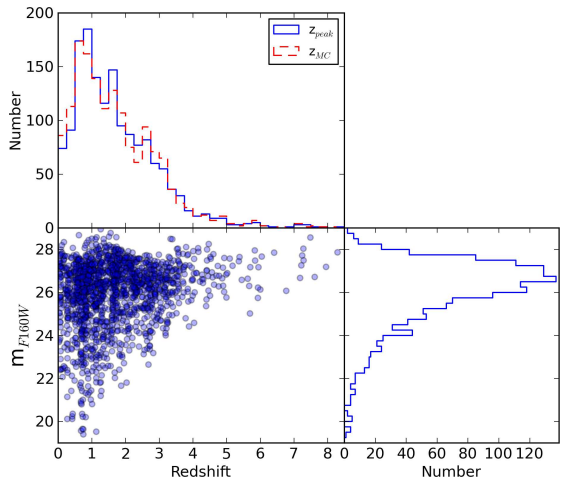}
\caption{The redshifts and total AB apparent magnitudes in the F160W filter for objects extracted from the HUDF.   A comparison of the distributions for z$_{peak}$ and z$_{MC}$ photometric redshift estimates from \textsc{EAZY} is provided in the top panel.  The histograms in the top and right panels are binned with a width of 0.25 in both redshift and magnitude. \label{fig:zdist}}
\end{center}
\end{figure}

\subsection{Photometric Redshift Estimates}

As the HUDF lacks a sufficiently large spectroscopic sample representative of the range of galaxies contained in the field, neural network redshift codes are poorly suited to our analysis.   Instead, we estimate photometric redshift using the \textsc{EAZY} code \citep{Brammer08} for all galaxies that have at least seven broad-band flux measurements.  \textsc{EAZY} operates by fitting a linear combination of template SEDs to each galaxy.  The template set, which is described in detail in  \citep{Brammer08}, contains seven SEDs spanning a broad range in galaxy spectral types.  The selection of these templates have been optimized to fit a wide range of galaxy properties, while minimizing degeneracies in color and redshift.  In addition, an $H_{160}$-band flux prior is applied to limit catastrophic failures from remaining degeneracies.  

For each object, \textsc{EAZY} produces a redshift probability distribution, which can be used to determine the photometric redshift in multiple ways.  The discrete peak of this distribution is returned as $z_{peak}$, and $z_{MC}$ provides a redshift from Monte Carlo sampling of the probability distribution.  For objects in which the redshift probability distribution is narrowly defined, the difference between these estimates is negligible.  However, in cases where the probability distribution is broad, $z_{MC}$ provides a more honest sampling within the error in the best-fit photometric redshift.  For comparison, we show the photometric redshift distributions of $z_{peak}$ and $z_{MC}$ for the 1553 galaxies in the HUDF in Figure~\ref{fig:zdist}.  While the results do slightly differ, a Kolmogorov-Smirnof test of the two distributions returns a statistic of 0.027, indicating that these differences may be considered insignificant for the average object in this sample.  A comparison of the relative differences of $z_{peak}$ and $z_{MC}$ estimates for objects matched to the FIREWORKS catalog implies that the difference between these estimates is negligible (differing on the 1\% level).  For the remainder of this work, we adopt $z_{peak}$ as the photometric redshift estimate of choice, since the $z_{MC}$ values are, by definition, impossible to exactly reproduce.

\begin{figure}[t!]
\centering
\epsscale{0.9}
\begin{center}
\plotone{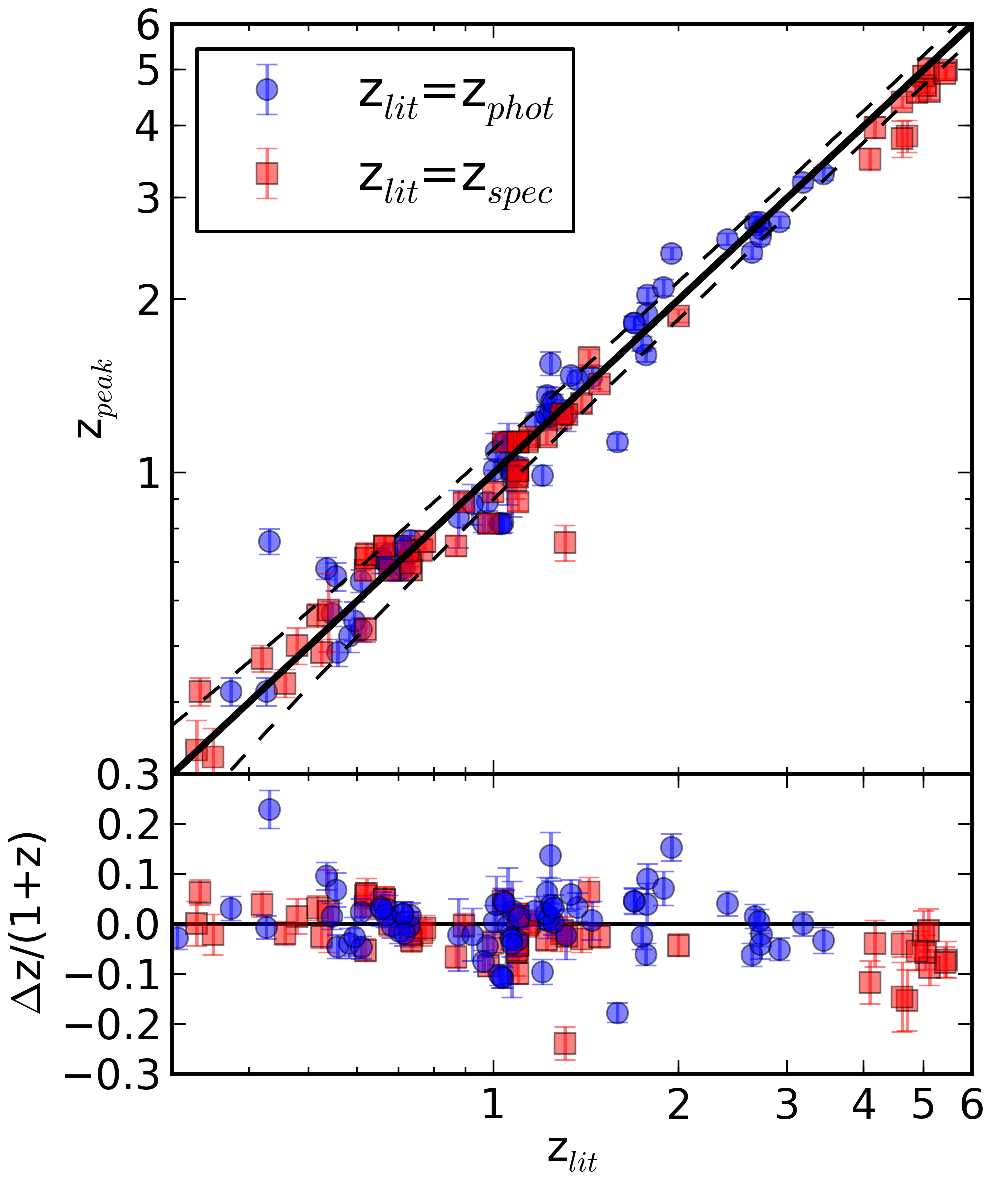}
\caption{A comparison of z$_{peak}$ photometric redshift estimates from \textsc{EAZY} with the published spectroscopic measurements for 63 objects from the GRAPES \citep{Pirzkal04} and PEARS \citep{Straughn08} catalogs (red).  The comparison to photometric redshift estimates for 82 unblended objects detected in the FIREWORKS catalog \citep{Wuyts08} are shown in blue.  Dashed lines indicate the scale of the normalized median absolute deviation of the differences between the photometric redshifts measured in this work and to values from the literature, $\sigma_{NMAD}=0.055$. \label{fig:photspec}}
\end{center}
\end{figure}

In order to test the accuracy of the $z_{peak}$ photometric redshifts, we cross-match our catalog with the \emph{FIREWORKS} catalog of \citet{Wuyts08}, the Grism ACS Program for Extragalactic Science (GRAPES; \citet{Pirzkal04}) and Probing Evolution and Reionization Spectroscopically (PEARS; \citet{Straughn08}) catalogs \citep{Rhoads09}.   Among the 1553 galaxies in this $H_{160}$-selected HUDF catalog,  63 have spectroscopic redshifts in the aforementioned datasets, extending to $z\sim5.5$.  The comparison of photometric and spectroscopic redshifts for the objects matching existing catalogs is shown in Figure~\ref{fig:photspec}.  The mean difference calculated for (z$_{spec}$-z$_{peak}$)/(1+z$_{spec}$) over the range available is 0.017.  We find a normalized median absolute deviation, $\sigma_{NMAD}$, of 0.055, as defined by:

\begin{equation}
\sigma_{NMAD} = 1.48 \times \mathrm{median} \Bigl\vert \Bigl( \frac{\Delta z - \mathrm{median} (\Delta z)}{1+z_{spec}}\Bigr) \Bigr\vert \label{eq:nmad}
\end{equation}

A total of 82 unblended objects in the $H_{160}$-selected galaxy catalog had photometric matches in the \emph{FIREWORKS} catalog.  A comparison of the photometric redshift estimates for these objects is also shown in Figure~\ref{fig:photspec}, with $\sigma_{NMAD}=0.048$.  The errors in the photometric redshifts we measure are comparable to measurements at $z<4$ in other deep fields \citep[$\sigma_{NMAD}=0.046$;][]{Brammer08}. 

For galaxies with $z>3$ the $\sigma_{NMAD}$ with respect to the spectroscopic redshifts nearly doubles to $\sigma_{NMAD}=0.095$, and the mean difference between spectroscopic and photometric redshifts increases to $(z_{spec}-z_{peak})/(1+z_{spec})=0.07$, indicating a potentially significant bias in the photometric estimates at higher redshifts.  The scale of this worsening accuracy with redshift is consistent with the commissioning tests of \citet{Brammer08}, though the performance of \textsc{EAZY} beyond $z=4$ has been limited to only a few galaxies.  We have examined the best-fit SEDs for these high-redshift sources and found that the worsening agreement of the photometric redshifts with the spectroscopic measurements at $z>3$ results from the difficulty of fitting the predominantly steep SEDs of young highly star forming galaxies at these redshifts.   While the trend worsens with redshift, the redshift bin sizes we choose for the analysis presented in Section 4 are sufficiently large ($1\leq\delta z\leq2$) to make these uncertainties negligible.


\subsection{Stellar Population Estimates}

Stellar population synthesis (SPS) models provide a means of estimating the evolution of an integrated galaxy spectrum, given an initial mass function (IMF), star formation history, and metallicity.  This technique, first employed by \citet{Tinsley68}, has shown dramatic improvement in recent years and is now a well-accepted method for deriving the estimated physical properties of galaxies from precisely measured SEDs. 

The galaxy properties output by SPS modeling are inherently dependent on the input parameters, such as the IMF.  While \citet{Chabrier03} and \citet{Salpeter55} produce similar colors for galaxies at $z>2$, \citet{Chabrier03} predicts stellar masses that are $\sim1.6$ times smaller.  The \citet{Chabrier03} IMF has been favored as of late and used in stellar mass estimates of high-redshift galaxies \citep[e.g.,][]{Papovich11}.  However, in a recent study of the stellar absorption features in local elliptical galaxies, \citet{PvDConroy10} revealed an abundance of low-mass stars, contributing to more than 60\% of the total stellar mass.  This finding suggests that the IMF of massive star-forming galaxies in the early Universe likely resembled a ``bottom-heavy" IMF, more akin to, but steeper than Salpeter.

 \begin{figure}[t!]
\centering
\begin{center}
\epsscale{1.0}
\plotone{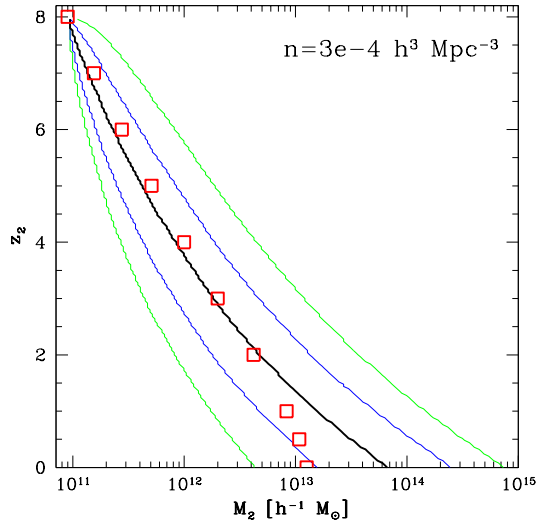}
\caption{The evolution of dark-matter halo mass and constant number density selection predicted from extended Press-Schechter modeling. Black line: median of the probability distribution that a $z_1=8$ halo with $M_1 = 10^{11} M_{\sun}$ evolves into a halo of mass $M_2$ at $z_2$ (from Trenti, Stiavelli \& Santos 2008).  Blue and green contours respectively enclose 68\% and 95\% of the halo distribution, which exhibits increasing scatter in mass with time. The open red squares show the mass of dark-matter halos with constant number density $n=3\times$10$^{-4}$ Mpc$^{-3}$, demonstrating that the selection employed in this paper is an effective tracer of halo growth over the full range of redshift we consider. \label{fig:eps_evol}}
\end{center}
\end{figure}


We fit stellar population models to the measured SEDs at the redshift determined by \textsc{EAZY} using the \textsc{FAST} algorithm described in \citet{Kriek09}.  We choose to employ a \citet{Salpeter55} IMF and the SPS models of \citet{bc03}, as these seem to be most effective at modeling massive star-forming galaxies at high redshift.  We also assume an exponentially declining ($\tau$ model) star formation history and solar metallicity.  In estimating the star formation rates, we apply the prescriptions of \citet{Wuyts11}, by requiring a minimum e-folding time of log($\tau_{min}$) $=8.5$, which has been shown to best reproduce the low-to-intermediate SFRs within the \citet{bc03} framework.  For the estimation of stellar masses we allow more freedom in the best-fit SPS models, setting log($\tau_{min}$) $=7$.  We then estimate the specific star formation rate of each object using the combination of the results from these methods.

For each object, \textsc{FAST} determines the best-fit to a six-dimensional cube of SEDs generated with a range in each of the following stellar population properties (age, star formation timescale, dust content, metallicity, and redshift), though we have fixed the redshift and metallicity in each case.  In addition to the best-fitting results, \textsc{FAST} computes 68\% confidence intervals on each stellar population parameter.  These are determined using Monte-Carlo simulations, which re-fit each observed SED 100 times after perturbing the flux measurements randomly within the photometric errors.   As a result we recover realistic random errors for each galaxy's estimated stellar mass, star formation rate, and age, with which we may now investigate their cosmologically-averaged evolution in the following section.

We note that we have required a fixed metallicity, which should be a poor assumption, particularly for low-mass galaxies at high redshifts.   Fitting the sample with metallicity as a free parameter indicates that a fixed solar metallicity will affect the best-fit estimated star formation rates of the galaxies as a function of redshift.  However, while this bias is clearly evident within the full photometric catalog, the galaxies above the lower mass limit of our analysis in the following section (log M$_{*}[M_{\odot}]>9.5$), show no significant departure from the results fit with a fixed solar metallicity.  Thus, the influence of evolving metallicity should not to inflict a substantial bias on the analysis presented in this work.




\section{Analysis}

\subsection{Constructing Samples with Constant Cumulative Number Density}

\begin{figure}[t!]
\centering
\begin{center}
\epsscale{1.25}
\plotone{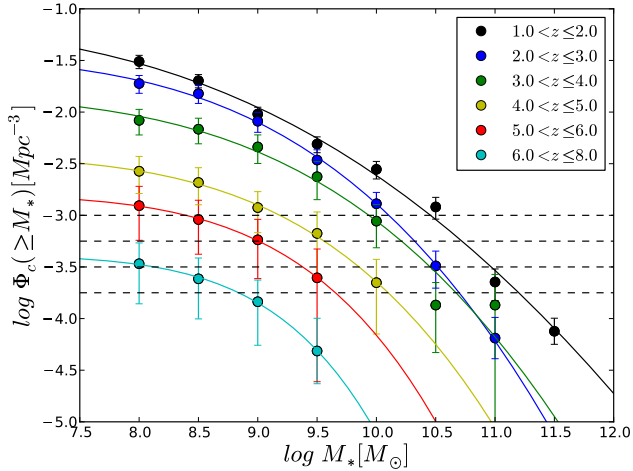}
\caption{The cumulative number density of galaxies in the HUDF as a function of minimum mass, calculated for a range of redshift bins.  Exponential fits to each are overplotted, which incorporate errors from cosmic variance that are estimated according to the prescriptions of \citet{Moster11}. Four lines of constant cumulative number density are included as black dashed lines, defining the subsamples of galaxies examined in this work. \label{fig:massfunc}}
\end{center}
\end{figure}

In order to best approximate the evolution of similar galaxies across a wide range in redshift, we have chosen to divide the galaxies into samples of constant cumulative number density, in a similar method to that utilized by \citet{CW09}, \citet{PvD2010}, \citet{Papovich11}.  Contrary to galaxies selected by parameters which strongly evolve with time, such as mass, color, or luminosity, number density-selected samples more closely track the cosmologically-averaged progenitors and descendants of galaxies with minimal bias over a range in redshift   This allows for the measurement of the average growth in stellar and gas mass, and average star formation rates for galaxies at a non-evolving number density over time.

As shown by \citet{PvD2010}, even in the presence of mergers, samples selected by this method are relatively pure tracers of galaxy evolution.  To precisely quantify the errors in cumulative number density selection caused by merger activity, \citet{Papovich11} applied this selection to the simulated merger trees produced by the Millenium Simulation \citep{Springel05b}.  Their results indicated that a constant cumulative number density selection recovers 60-80\% of halo descendants from their original progenitors over the redshift range $3<z<7$, a sample purity that is not possible by other observational methods of selection.

The purity of the sample can also verified by considering dark-matter halo evolution under extended Press-Schechter modeling \citep[e.g.,][]{Lacey1993}. This is illustrated in Figure~\ref{fig:eps_evol}.  For an halo of mass $M_1=10^{11}~M_{\sun}$ at $z_1=0$ (with number density $n=3\times$ 10$^{-4}$ Mpc$^{-3}$), we compute following \citet{Trenti08} the evolution of its descendant mass $M_2>M_1$ at $z_2<z_1$ (solid black line shows the median of the distribution, while blue and green lines are one and two sigma confidence intervals). The red squares in the figure show the mass at $z_2$ of dark-matter halos at fixed number density $n=3\times$ 10$^{-4}$ Mpc$^{-3}$, demonstrating that number density selected samples indeed trace the evolution of dark-matter halos very well down to $z\sim2$, and extending to a wider redshift range the validity of the \citet{Papovich11} findings. Interestingly, the number density selection appears to be less effective at $z\lesssim 2$, although it still captures the growth of dark-matter halos within 1$\sigma$.

In order to create samples of constant cumulative number density in the HUDF, we begin my measuring the cumulative mass function of galaxies, covering a wide range of redshifts and stellar masses, as shown in Figure~\ref{fig:massfunc}.  A compilation of these measurements is additionally provided in Table~\ref{tbl-2}.  Because of the exceptionally small size of the HUDF, cosmic variance contributes significantly to the error budget on this measurement.  Errors have therefore been calculated according to the prescriptions of \citet{Moster11}, to account for the fluctuations in the observed galaxy number counts that are expected as a function of mass for a field of this size. We fit an exponential curve to each slice in redshift, which model the data well.  The overlap of the curves tracing $2<z\leq3$ and $3<z\leq4$ in the high-mass regime (log M$_{*}[M_{\odot}]>10.5$) is likely one result of cosmic variance in this exceptionally small field.

\begin{figure}[t!]
\centering
\begin{center}
\epsscale{1.25}
\plotone{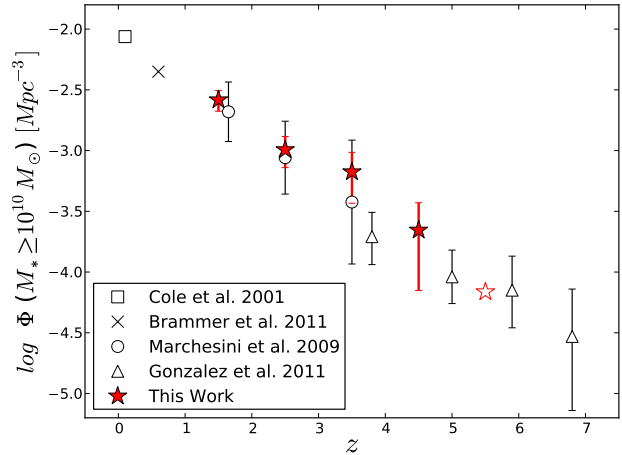}
\caption{The evolution of the number density of galaxies with log$(M_{*}[M_{\odot}])>$10 measured in this work, with measurements from various publications overplotted for comparison.  Published measurements that incorporate an alternate IMF have been corrected to Salpeter (+0.2 dex), and all are compared using a flat $\Lambda$--dominated CDM cosmology with $\Omega_m=0.3$, $H_0=70$ km s$^{-1} $Mpc$^{-1}$.  Presumably as a consequence of the small angular size of the HUDF, we detect no galaxies with log$(M_{*}[M_{\odot}])>10$ at $z>5$.  The unfilled red star marks the number density we expect at $z=5.5$, given the best fit to the cumulative mass function at this redshift in Figure ~\ref{fig:massfunc}.  \label{fig:ndensevol}}
\end{center}
\end{figure}

\begin{table*}
\centering
\begin{minipage}{0.95\textwidth}
\begin{center}
\caption{Cumulative Number Densities of Galaxies in the HUDF\label{tbl-2}}
\begin{tabular}{rrrrrrrr}
\tableline\tableline
z$_{min}$ & z$_{max}$   & $<$V$>$ & M$_{*}$ &  Number & log $\Phi_{c}$ & $\sigma_{-}$(log$\Phi_{c}$)  & $\sigma_{+}$(log$\Phi_{c}$) \\
& & (Mpc$^{3}$) & (M$_{\odot}$) & (M$>$M$_{*}$) &  (Mpc$^{-3}$) & & \\
\tableline
1.0 & 2.0 &  13269 & 8.0 & 410 & $-$1.510$^{a}$ & 0.069 & 0.060 \\
... & ...  & ... & 8.5 & 267  &  $-$1.696 & 0.069 &  0.060 \\
... & ... & ... & 9.0 & 127 & $-$2.019 &  0.078 & 0.066 \\
... & ... & ... & 9.5 & 65 &  $-$2.310 & 0.083 & 0.070 \\
... & ... & ... & 10.0 &  37 & $-$2.555 & 0.093 & 0.077 \\
... & ... & ... &  10.5 & 16 &  $-$2.919 & 0.118 & 0.093 \\
... & ... & ... &  11.0  & 3  &  $-$3.646 & 0.179 & 0.126 \\
... & ... & ... &  11.5  & 1  &  $-$4.125 & 0.126 & 0.126 \\
2.0 & 3.0 &  15445 &  8.0 & 292 & $-$1.723$^{a}$ & 0.095 & 0.078 \\
... & ... & ... &  8.5 & 233  & $-$1.821$^{a}$ & 0.095 & 0.078 \\
... & ... & ... &  9.0 & 126 &  $-$2.088 & 0.108 & 0.086 \\
... & ... & ... &  9.5 & 53 &  $-$2.464 & 0.129 & 0.099 \\
... & ... & ... & 10.0 & 20 & $-$2.888 & 0.147 & 0.110 \\
... & ... & ... & 10.5 & 5 & $-$3.489 & 0.215 & 0.143 \\
... & ... & ... & 11.0 & 1 & $-$4.187 & 0.200 & 0.200 \\
3.0 & 4.0 &  14797 & 8.0 & 123 &  $-$2.080$^{a}$ & 0.141 & 0.106 \\
... & ... & ... &  8.5 & 101 & $-$2.166$^{a}$ & 0.141 & 0.106 \\
... & ... & ... &  9.0 & 68 &$-$2.334$^{a}$ & 0.160 & 0.117 \\
... & ... & ... &  9.5 & 35 &  $-$2.626 & 0.222 & 0.146 \\
... & ... & ... & 10.0 & 13  & $-$3.056 & 0.258 & 0.161 \\
... & ... & ... & 10.5  & 2   &  $-$3.870 & 0.460 & 0.218 \\
... & ... & ... & 11.0  & 2  &  $-$3.870 & 1.706 & 0.297 \\
4.0 & 5.0  &  13466 &  8.0 & 36 & $-$2.573$^{a}$ & 0.216 & 0.144 \\
... & ... & ... &   8.5 & 28 &  $-$2.682$^{a}$ & 0.216 & 0.144 \\
... & ... & ... &   9.0 & 16 &$-$2.925$^{a}$ & 0.243 & 0.155 \\
... & ... & ... &   9.5 & 9 &  $-$3.175 & 0.412 & 0.208 \\
... & ... & ... &  10.0 & 3 & $-$3.652 & 0.497 & 0.226 \\
5.0 & 6.0 &  12090  & 8.0 & 15 & $-$2.906$^{a}$ & 0.336 & 0.187 \\
... & ... & ... & 8.5 & 11  & $-$3.041$^{a}$ & 0.336 & 0.187 \\
... & ... & ... & 9.0  & 7 & $-$3.237$^{a}$ & 0.375 & 0.198 \\
... & ... & ... &  9.5 & 3 & $-$3.605 & 1.006 & 0.279 \\
6.0 & 8.0 &    20604 &   8.0 &  7 & $-$3.469$^{a}$ & 0.388 & 0.202 \\
... & ... & ... &    8.5  & 5  & $-$3.614$^{a}$ & 0.388 & 0.202 \\
... & ... & ... &    9.0  & 3  & $-$3.836$^{a}$ & 0.422 & 0.210 \\
... & ... & ... &    9.5  & 1  & $-$4.310$^{a}$ & 0.317 & 0.317 \\
\tableline
\end{tabular}
\footnotetext[1]{Measurement significantly affected by incompleteness.} 		
\end{center}
\end{minipage}
\end{table*}



\begin{figure*}[t!]
\epsscale{0.82}
\begin{center}
\plotone{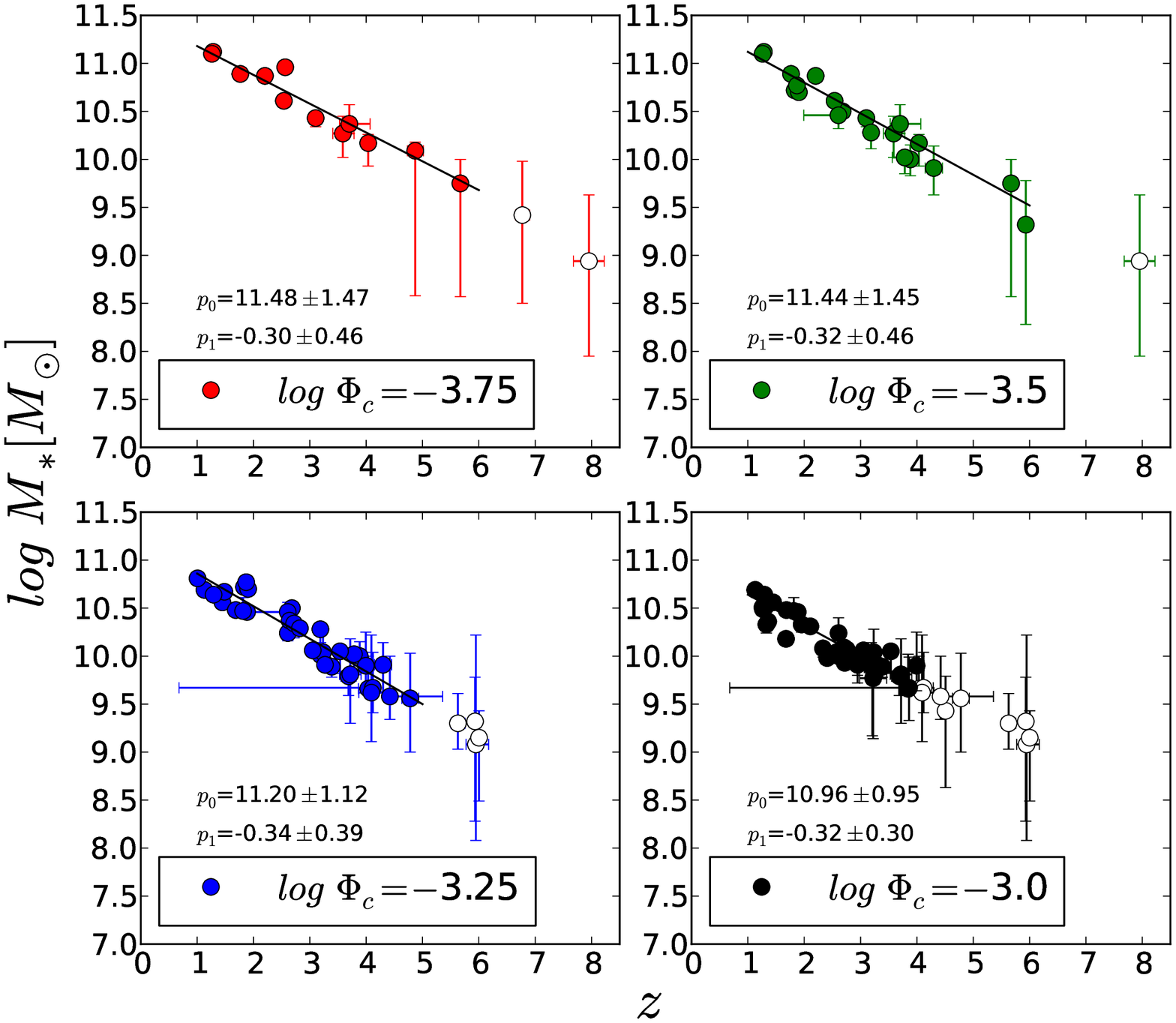}
\caption{Measurements showing the evolution of the stellar mass for four samples of galaxies selected at different values of constant cumulative number density.  Open circles represent measurements in regions of incompleteness, as described in the text.  Minimized $\chi^{2}$ fits to the moving mean of the data of the form: log M$_{*}$ (M$_{\odot}$) = $p_{0}$+$p_{1}$z are overplotted where the data are complete, and the parameters of each fit are provided.   \label{fig:massevol_median}}
\end{center}
\end{figure*}

In Figure \ref{fig:ndensevol} we compare our measurements of the evolving number density of galaxies with log$M_{*}[M_{\odot}]>$10 with published values from the literature.  We find excellent agreement with the results of \citet{Marchesini09} and \citet{Gonzalez11}, whose measurements overlap in redshift with the available range of our sample.  While the measurements of \citet{Marchesini09} have been calculated assuming a Kroupa IMF (unlike those of \citet{Gonzalez11}, which assume a Salpeter IMF), the errors have been calculated so as to accommodate variance due to this difference.  Thus, these measurements are directly comparable and are fully consistent within the quoted errors.  Literature data presented in this plot have likewise been corrected to a Salpeter IMF and compared using the most commonly applied cosmology (flat, $\Lambda$-dominated CDM with $h_{0}=0.7$, $\Omega_{m}=0.3$).

\begin{figure}[t!]
\epsscale{1.25}
\begin{center}
\plotone{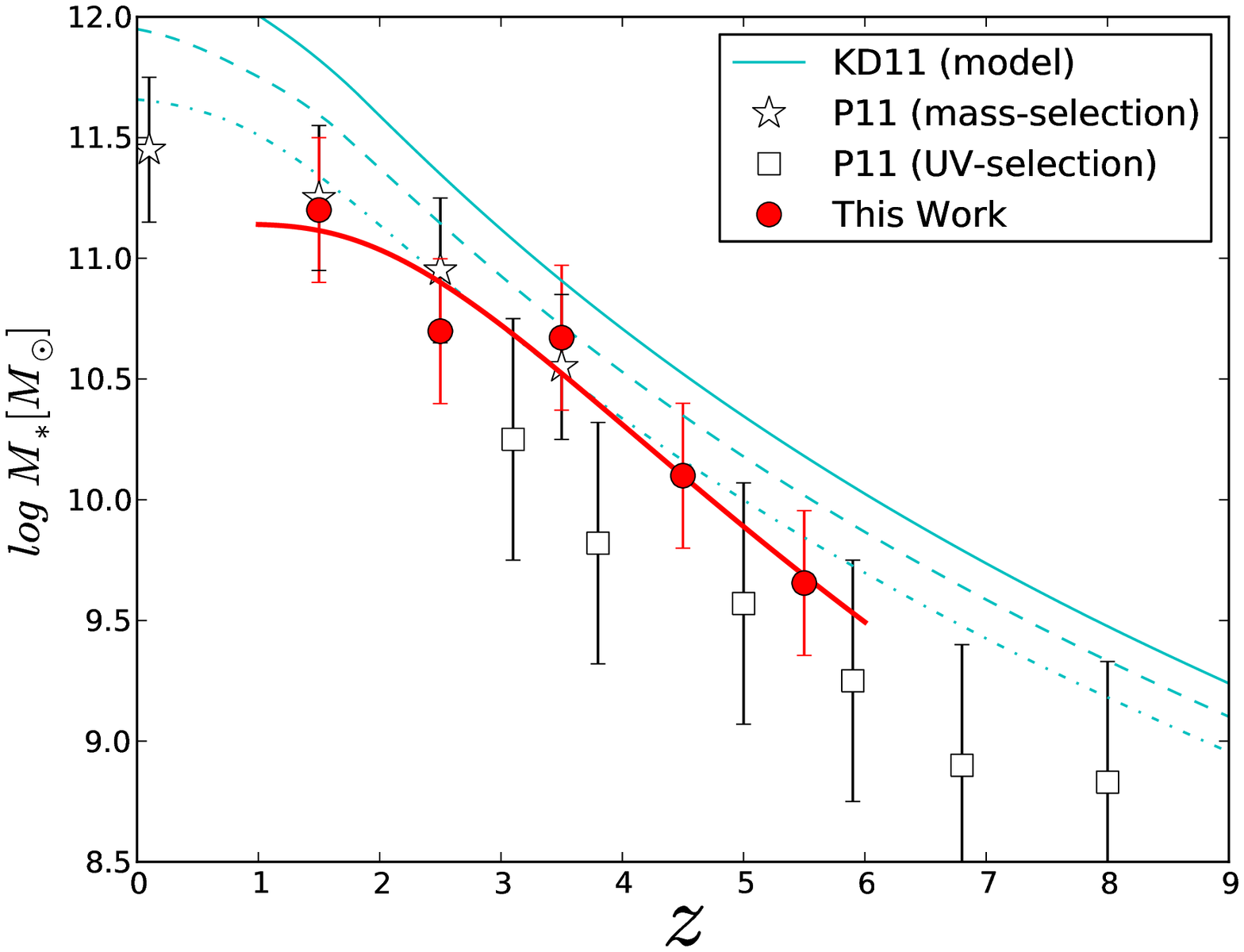}
\caption{The mean stellar mass growth inferred from the HUDF cumulative mass function at log $\Phi_{c}=-$3.75 Mpc$^{-3}$, shown as a function of redshift where complete.  A minimized $\chi^{2}$ exponential fit to the data is overplotted in red.  For comparison, measurements from \citet{Papovich11} for similar cumulative number density, log $\Phi_{c}=-$3.7 Mpc$^{-3}$, are shown in black.  Stars provide measurements determined using the integrated mass function from the mass-selected sample of \citet{Marchesini09}; squares indicate measurements inferred from the integrated UV luminosity functions of  \citet{RS09}, \citet{Bouwens07}, and \citet{Bouwens11a}.  These literature values have been corrected to Salpeter (+0.2 dex), in order to facilitate a direct comparison. Model curves from \citet{KD11} are shown in cyan, with contours indicating, from top to bottom, the mass growth of galaxies with log $\Phi_{c}=-$3.7 Mpc$^{-3}$ selected with 100\%, 50\% and 25\% completeness. \label{fig:massfits}}
\end{center}
\end{figure}

We define samples with constant cumulative number density (hereafter, $\Phi_{c}$) for each redshift range by determining the stellar mass at the intersection of the exponential fits in Figure~\ref{fig:massfunc} and each of the horizontal dotted lines, equivalent to $\Phi_{c}=-3.0$, $-$3.25, $-$3.5, and $-$3.75 Mpc$^{-3}$.  This provides the typical masses of galaxies and their progenitors at a constant number density over the range of observable redshifts.   For each of the four number density-selected samples, we fit a linear function to the evolving log $M_{*}[M_{\odot}]$ at constant number density, as a function of redshift.  We then select all galaxies in the photometric catalog with estimated stellar masses within 0.2 dex of the best fit linear function at any given redshift.  

For each of the chosen levels of $\Phi_{c}$, in order of increasing number density, this method selects samples with total sizes of 14, 21, 40, and 53 galaxies, which will be later split into smaller bins in redshift for further analysis of their redshift evolution.  Despite the unsurpassed depth of the WFC3/IR imaging, the small area of the HUDF field limits the number of samples with constant $\Phi_{c}$ that are fully accessible across the redshift range of $1<z<6$.  However, such measurements are possible for the galaxies with log $\Phi_{c}\leq-3.5$ Mpc$^{-3}$.   

From an examination of the measured signal-to-noise of the IRAC and H$_{160}$ fluxes as a function of galaxy redshift and estimated stellar mass, we find these data are approximately complete for log M$_{*}[M_{\odot}]>9.5$ at $z=5$.  Thus, although the catalog includes galaxies at redshifts extending to $z\sim8$, none of our number density-selected samples are complete beyond $z\sim6$.   Galaxies selected at log $\Phi_{c}\leq-3.5$ are complete to a limit of $z\sim6$, but examining galaxies at progressively higher cumulative number densities requires a progressively lower mass detection threshold.  Thus, our mass limit translates into progressively lower redshift limits for galaxies with increasingly higher cumulative number densities:  galaxies selected at log $\Phi_{c}=-3.25$ Mpc$^{-3}$ are complete to $z\sim5$, and those at log $\Phi_{c}=-3.0$ Mpc$^{-3}$ are complete only to $z\sim4$.


\subsection{Extending the analysis to $z\sim0.5$, with the FIREWORKS Catalog}

Due to the exceptionally small area covered by the HUDF, our measurements become increasingly susceptible to cosmic variance for the most massive and lowest redshift galaxies.  To circumvent this problem at low redshift, we have incorporated data from the $K_{s}$ -selected FIREWORKS catalog of \citet{Wuyts08} in order to provide a continuing measurement of the evolution of these number density selected galaxy samples down to $z\sim0.5$.  Though limited to lower redshifts, the FIREWORKS catalog spans a significantly larger region of overlapping sky, equal to 138 arcmin$^{2}$, and its included SEDs span an even broader range in wavelength coverage, extending from the $U$-band to 24$\mu m$ MIPS imaging.   

We calculate the photometric redshifts and stellar population estimates for galaxies in the FIREWORKS catalog with $\geq$30\% of maximum coverage in all observed bands, using the same  \textsc{EAZY} and \textsc{FAST} prescriptions described in the previous sections.  For objects included in both the $H_{160}$-selected catalog and FIREWORKS, comparisons of the photometric redshifts we determine with those provided by \citet{Wuyts08} are shown in Figure~\ref{fig:photspec}.   The photometric redshifts show close agreement, with little evidence for catastrophic failures and $\sigma_{NMAD}=0.055$.

\section{Evolution at Constant Number Density}
\subsection{Stellar Mass Evolution}

In Figure~\ref{fig:massevol_median}, we present the best-fit stellar masses of the individual $H_{160}$-selected galaxies as a function of redshift for four samples selected at different constant cumulative number densities.  Linear fits to the moving average in each of the four samples are overplotted where the data are complete.  We note that the data closely follow a linear relation in these plots as a direct result of the method of selection described in the previous section.

The amplitude of each fit to the mass evolution at constant number density in Figure~\ref{fig:massevol_median} rises with decreasing number density, as expected with this method of selection, indicating that samples at an increasing number density trace galaxies with decreasing mean stellar mass.  The best-fits to the slopes of the mass evolution do not significantly vary over the range in number density we examine, implying that the mass evolution occurs with approximately the same rate for each of these number density-selected samples.  The result is an exponential buildup equivalent to $\sim1.5$ orders of magnitude in stellar mass from $z\sim6$ to $z\sim2$, in each sample.  These observations are consistent with well-established evidence showing that massive galaxies build up the bulk of their stellar mass prior to $z=2$ \citep[e.g.,][]{Thomas05}.  The rate of growth we measure in the range $2<z<6$ also precisely agrees with predictions for the mass evolution of star-forming galaxies from simulations \citep{OD2008}.

In Figure~\ref{fig:massfits} we compare our results with the measurements of \citet{Papovich11}, which incorporated two separate methods for inferring the mass growth of galaxies with log $\Phi_{c}=-3.7$ Mpc$^{-3}$.   The star-shaped points indicate measurements obtained by using the integrated mass function of \citet{Marchesini09} to determine the limiting mass of a sample with their desired cumulative number density.  The square points indicate measurements of  \citet{Papovich11} that were indirectly determined using the integrated combined UV luminosity functions of \citet{RS09}, \citet{Bouwens07}, and \citet{Bouwens11a}.  Each of these measurements has been corrected to match estimates from a Salpeter IMF (+0.2 dex).  

The slope of the mass growth we measure with time agrees with the mass-selected measurements of \citet{Papovich11} at the 1$\sigma$ level.  However, the amplitude of the fit we recover at similar number density  (log $\Phi_{c}=-3.75$ Mpc$^{-3}$) is consistently a factor of $\sim2$ higher in stellar mass than the UV-selected measurements over the overlapping redshift range.   The method of rest-frame UV-selection employed by \citet{Papovich11} in the higher redshift regime may account for this difference, since such a selection is likely to be less inclusive of galaxies with lower star formation rates.  

Recent simulations by \citet{KD11} suggest that the mean mass estimates of UV-selected galaxies from \citet{Papovich11} may suffer from $>75$\% incompleteness in their sampling of star-forming central galaxies in the high redshift regime.  This incompleteness is suspected to be due to suppressed star formation scattering galaxies below the detection threshold at any given mass.  As 45\% of K-selected galaxies have been found to exhibit suppressed star formation at $z=2$ \citep{Kriek06}, \citet{KD11} predict a $\sim50$\% incompleteness in mass-selected samples at high-redshift.   

For comparison to our measurements, we overplot the theoretical mass growth for a fully complete sample from \citet{KD11} as a solid cyan line in Figure~\ref{fig:massfits}.  The two dashed lines with progressively lower amplitudes give the expected growth curves of \citet{KD11} for samples with 50\% and 75\% incompleteness, respectively.  The mass evolution for a 50\% complete number density-selected galaxy sample estimated by \citet{KD11} agrees with our observations to within 1$\sigma$.   Thus star formation duty cycle effects could yet be contributing to some degree of incompleteness in our sample, despite the apparent improvement over the measurements of \citet{Papovich11} that we achieve by detecting in H$_{160}$ and incorporating SED constraints at 3.6 and 4.5 microns.   The fact that our data exhibit a closer agreement with both observational analyses, as compared to the predictions of \citet{KD11}, may also imply that the models require additional complexity to reproduce the results from observations.

The cosmological hydro-simulations of \citet{Jaacks12} have also suggested that he duty cycle of bursty star formation histories may result in incompleteness in high-redshift observations.  However, their simulations imply that this effect only significantly reduces the number of galaxies above the detection threshold of the HUDF  for stellar masses below log $M_{*}[M_{\odot}] = 9.0$.  As the mass-completeness threshold of our analysis (log $M_{*}[M_{\odot}]> 9.5$) is above this lower limit, it is not clear that bursty star formation histories alone can account for the difference in the number of galaxies observed, relative to the model expectations from \citet{KD11}.

\begin{figure}[t!]
\centering
\begin{center}
\epsscale{1.25}
\plotone{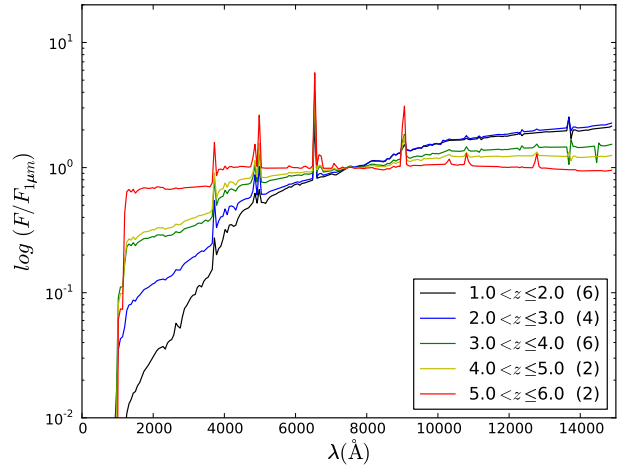}
\caption{The mean evolution of the rest-frame composite best-fit template SEDs for the sample of galaxies selected to have log $\Phi_{c}=-3.5$  Mpc$^{-3}$.  Each best-fit template SED has been normalized at 0.75$\mu m$ prior to stacking.  The number of SEDs contributing to each stack are provided next to each redshift range in the legend.   \label{fig:specevol}}
\end{center}
\end{figure}

\begin{figure*}
\epsscale{1.0}
\begin{center}
\plotone{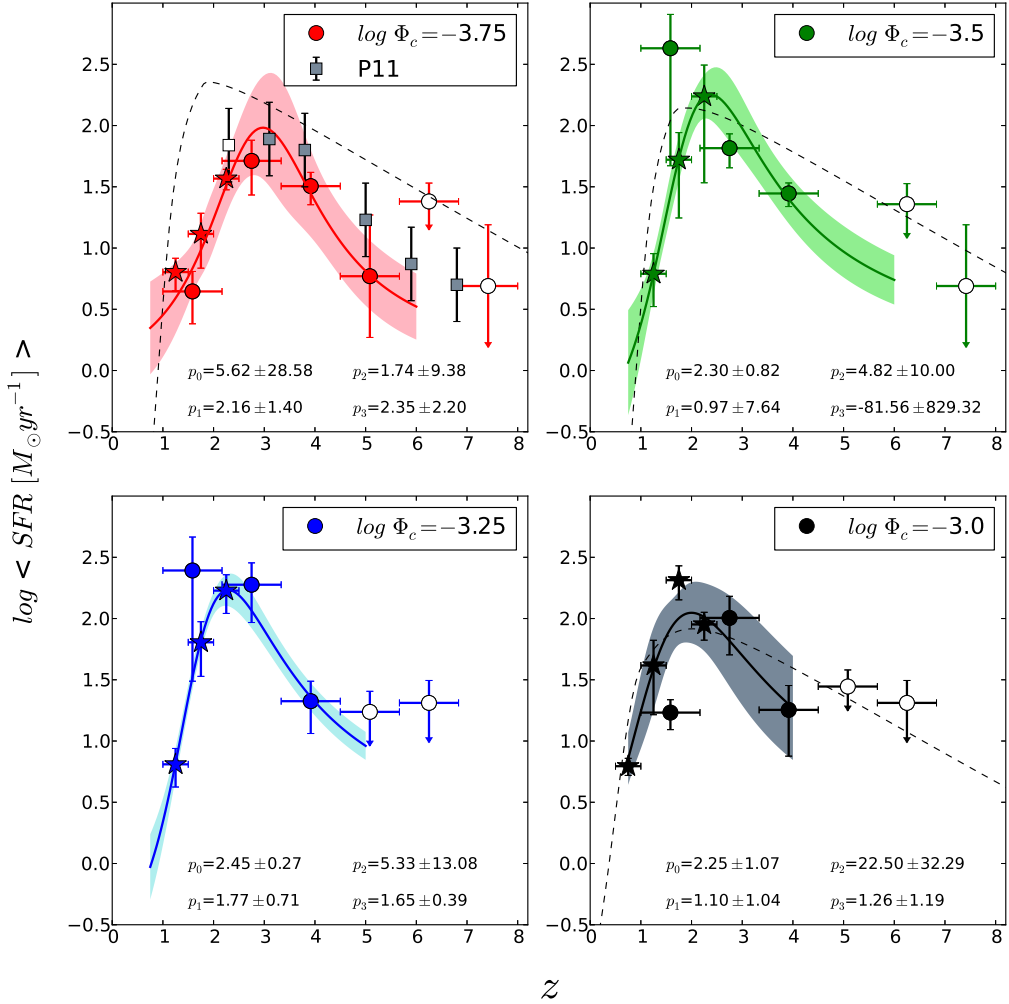}
\caption{The evolution of the mean star formation rates for four samples of galaxies with constant cumulative number density. Measurements presented in this work are plotted as circles, filled where the data are complete; lower redshift measurements marked with stars are derived using the FIREWORKS catalog of \citet{Wuyts08}. A best-fit asymmetric Lorentzian profile is fit to each sample at constant cumulative number density, and shaded regions indicate 68\% confidence intervals derived from the fitting.  Best fit parameters describing the fitting function (Equation~\ref{eq:lorentz}) are provided in each frame.  The fits indicate a steady increase in mean star-formation rates from $z\sim6$ to $z\sim2$, which turns over at progressively later times for samples with progressively lower cumulative number densities. Dashed curves in three of the four panels indicate the model predictions from \citet{KD11} for galaxies with the closest matching number densities: log $\Phi_{c}=-3.7$, $-$3.4 and $-$3.1 Mpc$^{-3}$.  In the lowest number density panel, measurements from \citet{Papovich11} with log $\Phi_{c}=-3.7$ are overplotted after correcting to a Salpeter IMF with filled squares, where deemed complete. \label{fig:sfrevol}}
\end{center}
\end{figure*}


\subsection{Star Formation Rate Evolution}


The evolution of the stellar masses selected at constant cumulative number density appears fairly smooth, largely as a consequence of the method of selection.  However, the star formation properties of the same galaxies exhibit greater scatter, relative to that of their best-fit stellar masses, when selected by the same method.  Thus, in order to examine the typical evolution in star formation rates with better statistical certainty, in the following section we further bin the data in redshift to examine the mean evolution in SFR among the four sub-samples previously chosen.

In Figure \ref{fig:specevol} we present the redshift evolution of the averaged best-fit template SEDs for galaxies in our sample, selected to have log $\Phi_{c}=-$3.5 Mpc$^{-3}$.  Each composite spectrum contains between 2 and 6 galaxies, the best-fit template SEDs of which have been shifted into the galaxy rest-frame and normalized at 0.75$\mu m$. The evolution of the composite spectral shape over the redshift range $1<z<6$ is consistent with an aging stellar population, providing a qualitative proof-of-concept for the method of cumulative number density selection and the usefulness of examining the mean evolution of stellar populations in galaxies selected by this method.

In Figure~\ref{fig:sfrevol} we present results for the cosmologically-averaged evolution of star formation rates in bins of $\delta z=1.25$ for the four samples of galaxies, defined by cumulative number densities of log $\Phi_{c}=-3.0$, $-$3.25, $-$3.5 and $-$3.75 Mpc$^{-3}$.   At redshifts where our galaxy samples are incomplete we plot the mean SFR with empty circles with errors to indicate that these measurements should be regarded as upper limits.  Incorporating measurements from the FIREWORKS catalog enables higher resolution of the peak SFR in each sample as well as the SFR evolution at $z<2$. 

For each of the galaxy samples, we observe a steady increase in star formation rate from $z\sim6$, which turns over around $2\lesssim z\lesssim3$.  This shape is well approximated by an asymmetric Lorentzian function, defined by:
\begin{equation}
y=\frac{p_{0}p_{1}}{2}\times\frac{\frac{p_{1}}{2}+2p_{2}(x-p_{3})}{(x-p_{3})^{2}+(\frac{p_{0}}{2})^{2}} \label{eq:lorentz}
\end{equation}
We fit each of the four samples with this profile and shade the regions defined by a 68\% confidence intervals in Figure~\ref{fig:sfrevol}.  We stress that this shape has no obvious physical interpretation, particularly given the logarithmic nature of the plot.  However, the $\chi^{2}$ values returned by fitting the data to this form are approximately unity, indicating that this profile may be an acceptable function for interpolating through the points of measurement.   While previous studies have chosen to fit a power law to the rising star formation rate (SFR) over this redshift range, we find that the asymmetric Lorentzian function better models the more gradual rise of the data, as well as providing a smooth transition to the downturn below $z\sim2$.

\begin{figure*}[t!]
\epsscale{1.0}
\begin{center}
\plotone{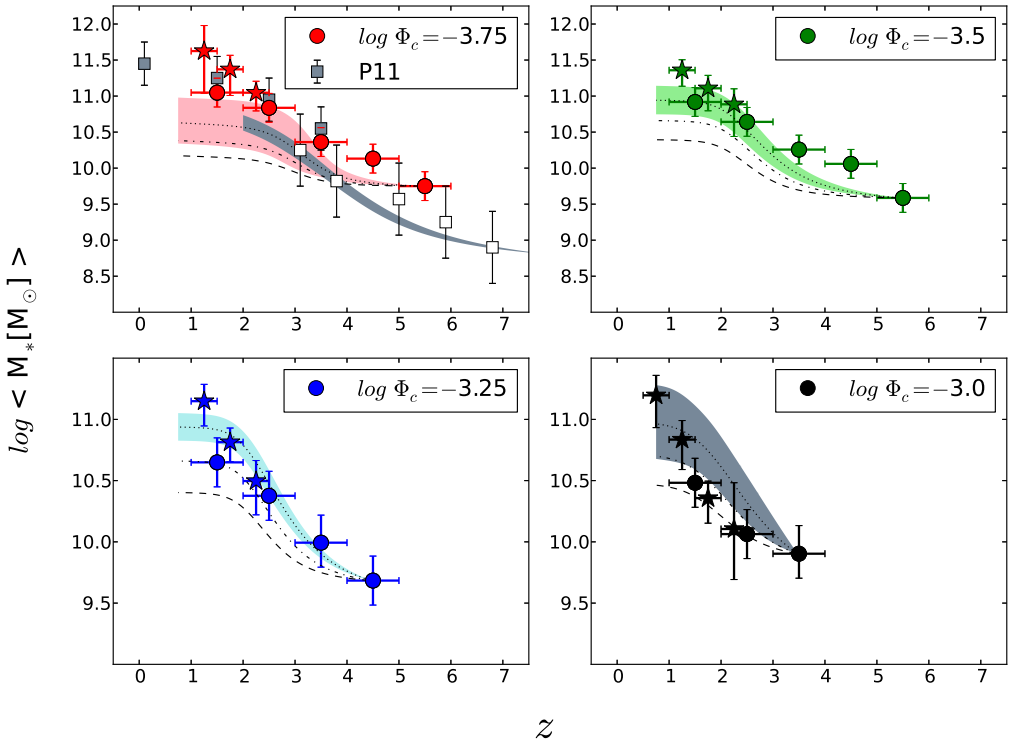}
\caption{Measurements of the mean stellar mass evolution for samples selected with various cumulative number densities.  As in previous figures, measurements from the HUDF catalog are shown as circles, and measurements from the FIREWORKS data are shown as stars. The predicted mass growth from the integrated observed star formation rate evolution at each number density is overplotted, using the best-fits to the SFR evolution in Figure \ref{fig:sfrevol} and assuming star formation duty cycles of 25\%, 50\%, and 100\% (dashed black lines, from bottom to top).  In each panel, shaded regions enclosing the 100\% duty cycle curve illustrate the 1$\sigma$ uncertainty in the predicted mass growth, which has been inferred from the fits to the observed SFR evolution. Measurements of the mass growth at log $\Phi_{c}=-$3.7 Mpc$^{-3}$ from \citet{Papovich11} are overplotted for comparison with our lowest number density sample; filled squares represent mass-selected measurements and open squares represent measurements inferred from the rest-frame UV luminosity function.  The grey curve underlying those points illustrates the stellar mass growth inferred from the \citet{Papovich11} SFR evolution measurements, when modeled using our chosen functional form (Equation 5). \label{fig:massgrowth}}
\end{center}
\end{figure*}

The fits to the data exhibit no significant difference in the slopes of the SFR evolution in the range $2<z<6$ among each of the samples selected at different constant number densities.  Our fits also suggest that all four samples achieve a similar maximum SFR (log SFR [M$_{\odot}$ yr$^{-1}$] $\sim2$) at $2\lesssim z\lesssim3$.  We find that galaxies with log $\Phi_{c}=-3.75$ Mpc$^{-3}$ reach a peak mean SFR at $z\sim3$; galaxies with $-3.5\leq$ log $\Phi_{c} \leq-$3.25 Mpc$^{-3}$ peak at $z\sim2.3$; galaxies with log $\Phi_{c}=-$3.0 Mpc$^{-3}$ peak at $z\sim2$.  These results indicate that the timing of the peak SFR is correlated with the cumulative number density, such that samples with lower cumulative number density (and thus, higher average stellar mass) reach a peak SFR at progressively earlier times.  This behavior appears consistent with what one would expect from cosmic downsizing \citep{Cowie96}, in which the epoch of major mass buildup in galaxies moves to lower redshift for increasingly lower-mass galaxies.
 
For comparison, we have overplotted measurements of the SFR evolution from \citet{Papovich11} in Figure \ref{fig:sfrevol}.  Although the mean SFR estimates from \citet{Papovich11} are approximately 2 times higher than what we measure at $z>3$, they still agree within the stated 1$\sigma$ errors. The evolution we measure in the SFR from $2<z<6$ is thus consistent with \citet{Papovich11}, despite the slight inconsistencies we find in the stellar mass evolution for the same cumulative number density (see also the comparison in \citet{Smit2012}).  This agreement in the measured SFR evolution is unsurprising, given that our rest-frame UV measurements are derived from overlapping datasets.

This rise in SFR with time, which we observe in each of the number density-selected samples prior to $z\sim3$, is predicted by theoretical models and hydro-simulations \citep[e.g.,][]{Robertson04, DeLucia06, Finlator06, FDO07, Brooks09, Trenti10, Finlator11, KD11}.  The number density-dependent evolution we measure in the cosmically-averaged SFRs is also consistent with the recent simulations of \citet{KD11}, who incorporate the effects of an evolving metallicity in modeling the star formation histories of galaxies.  In Figure \ref{fig:sfrevol} we overplot the predictions of \citet{KD11} for galaxies selected over a similar range in cumulative number density.  The agreement is impressive, especially for galaxies with log $\Phi_{c}\geq-3.5$ Mpc$^{-3}$, where our data are well-sampled.  The fact that the models closely track even the measurements at $z>5$ suggests that the mean star formation rates we measure in the extreme high-redshift regime are close to the predicted cosmological average, despite our possible mass incompleteness at those redshifts due to limitations from our photometric wavelength coverage and IRAC imaging depth. It is difficult to determine whether the departure of our measurements from the models of \citet{KD11} in the lowest number density sample (log $\Phi_{c}=-3.75$ Mpc$^{-3}$) is physical, as it could also be explained by the small number of galaxies included in the measurement, which will be most dramatically affected by cosmic variance.

\begin{figure*}[t!]
\epsscale{1.0}
\begin{center}
\plotone{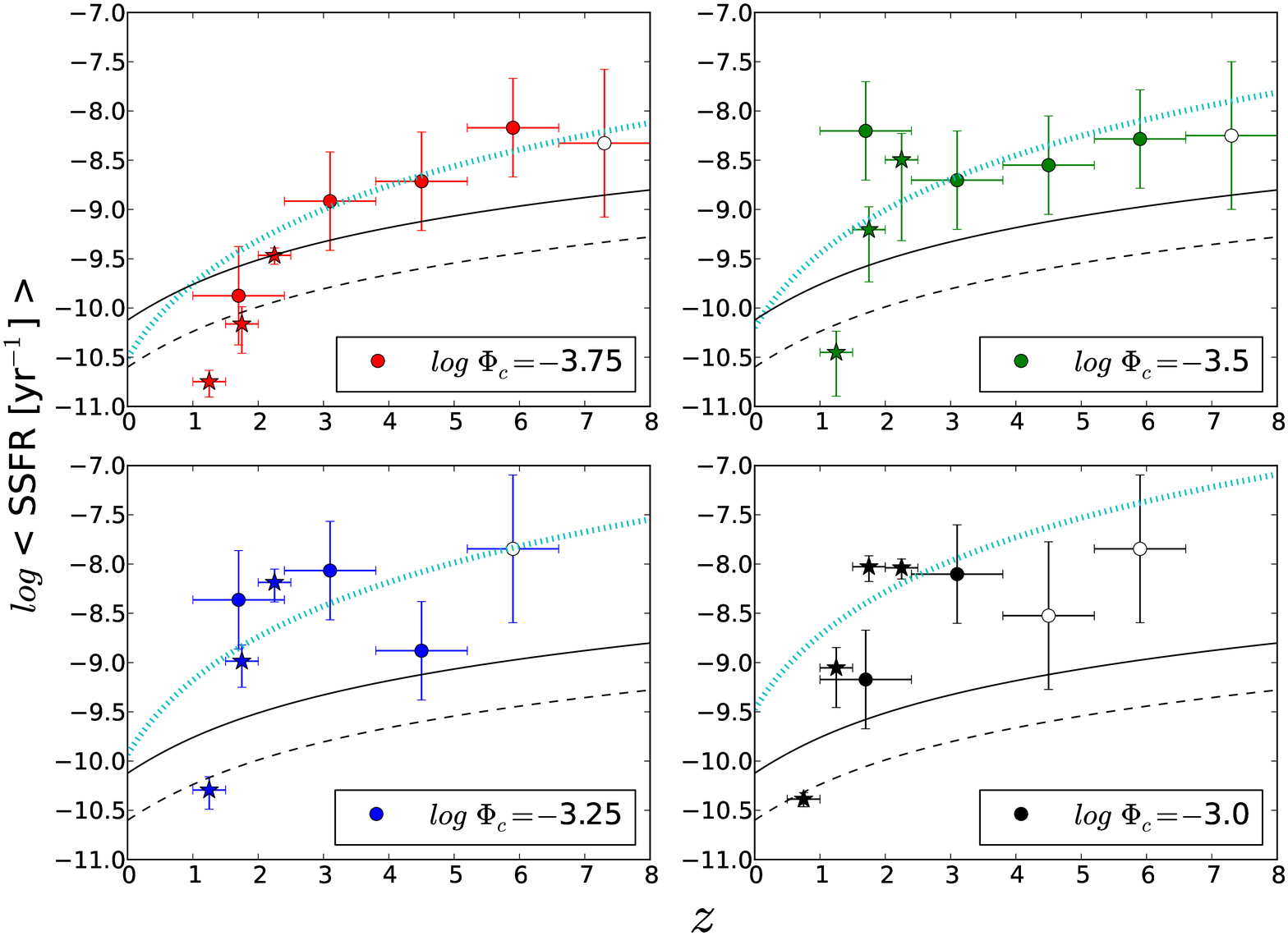}
\caption{The evolution of the average specific star formation rates for the four samples of galaxies selected at constant cumulative number density. As in the previous figures, measurements using the $H_{160}$-selected catalog are plotted as circles;  lower redshift measurements marked with stars are derived using the FIREWORKS catalog of \citet{Wuyts08}. The solid black line traces the inverse Hubble time ($1/t_{H}$ in $yr^{-1}$); the dashed black line, ($1/3t_{H}$).  The dashed cyan line provides the specific gas accretion rate predicted by \citet{ND08}, which scales as $(1+z)^{2.5}$ and has been normalized to best fit our measurements in the redshift range $2<z<6$. \label{fig:ssfrevol}}
\end{center}
\end{figure*}

\subsection{Mechanisms of Mass Build-Up}

With the stellar mass and star formation rate evolution each separately measured for the samples we have selected, we now consider how well the measured star formation rates alone can account for the stellar mass build-up we observe.  In Figure~\ref{fig:massgrowth}, we use the best fits to the evolution of the cosmologically averaged star formation rates to predict the stellar mass growth for the same samples of galaxies. In each panel, three tracks of mass growth derived from the integrated best-fit star formation rate evolution with respective duty cycles of 25\%, 50\%, and 100\% are plotted alongside the separately measured mean stellar mass as a function of redshift.  For each mass growth track with a duty cycle of 100\% we also provide the 68\% confidence interval.  In each case we assume a fractional mass loss of 30\% from star formation-driven winds, as traditionally applied to stellar populations modeled with a Salpeter IMF.  We note that the true fractional mass loss during star formation will depend on a number of factors, such as the choice of initial mass function and recycling timescale.  

At all number densities we examine, the mass growth inferred from the integrated SFR evolution broadly agrees with the observed mass evolution at $z\sim2$.  However, at lower redshifts, we find that the highest mass galaxies in our sample appear to grow more massive than what is suggested from their observed SFRs alone. For galaxies at the lowest number density we examine, which are equivalent to galaxies with log(M[M$_{*}$]) $\sim11.5$ at $z=0$, we find that even a duty cycle of 100\% under-predicts the $z=1.5$ stellar masses by more than a factor of 3 with $>68$\% confidence.  The inability of star formation alone to produce the observed buildup in stellar mass in this high-mass sample indicates that additional processes are responsible for contributing the residual mass growth, particularly at late times ($z<2$).  Mergers are the most likely mechanism for making up this difference, although obscured star formation could also play a substantial role.

Among the galaxies selected at progressively higher number densities, we find an increasingly better agreement between the mass growth expected from the evolution of the observed star formation rates within a reasonable range of duty cycles.   The stellar mass build up observed for the samples with log $\Phi_{c}=-3.0$ Mpc$^{-3}$ can be fully accounted for until $z\sim1.5$ with a duty cycle as low as 25\%.  Galaxies with cumulative number densities in the range $-3.5\leq$ log $\Phi_{c}\leq-3.25$ Mpc$^{-3}$ are also found to agree with mass build-up from star-formation alone until $z\sim2$ assuming duty cycles in the range 50$-$100\%.  

In general, at lower cumulative number densities (and thus, higher mean stellar mass), the mass growth from star formation alone falls increasingly further short of the observed stellar mass measurements.  
These data indicate that the fraction of mass build-up that may be accounted for from star formation with a constant duty cycle scales with cumulative number density, such that galaxies with higher number density build up more of their mass through star-formation alone, compared to their counterparts at lower number densities.  Considering that the mean stellar mass of a galaxy is inversely related to its cumulative number density, it may not be surprising that the mechanism of stellar mass build-up is dependent on the number density.  Lower-mass galaxies will have smaller gravitational potentials from which to attract mergers and recapture ejected mass from feedback processes.  The fact that the gap between the observed stellar mass and that which is expected from the observed star formation rates becomes especially prominent at $z<2$ suggests that mergers contribute significantly to the late-time mass growth in the galaxies with log $\Phi_{c}\leq-$3.5 Mpc$^{-3}$.  This finding is consistent with \citet{PvD2010}, who determined that in-situ star formation can only account for a small fraction ($\sim20$\%) of the mass growth of galaxies with similar mean mass at $z<2$.  

At face value, our results may appear to conflict with the findings of \citet{Papovich11}, in which the observed star formation rates were found to fully account for the average mass buildup of galaxies with even slightly lower number densities than those examined here (log $\Phi_{c}=-3.7$ Mpc$^{-3}$).  We note that the gap we find between the observed stellar mass growth and the growth inferred from SFR only becomes significant at $z<2$, the upper limit of the primary redshift range examined by \citet{Papovich11}.   However, the fact that \citet{Papovich11} SFR evolution measurements were fit with an exponential function, which continually increases with time, suggests that the  difference in fitting models could intrinsically produce greater build up of mass compared to ours in the low-redshift regime ($z<3$), where the SFRs actually appear to decline.   

In order to examine the effect of the chosen SFR evolution fitting function on the implied stellar mass growth, in Figure 11 we directly compare the stellar mass growth inferred from the \citet{Papovich11} SFR evolution and stellar mass estimates,after fitting with the same functional form applied to our data and assuming a 100\% duty cycle.    In agreement with the results of \citet{Papovich11}, we find that the observed SFR evolution can fully account for the stellar mass growth, as estimated by \citet{Papovich11} to $z\sim3$.  At lower redshifts their SFR evolution data become incomplete, but the trend of inferred mass growth suggests that the observed stellar mass growth begins to exceed the implied growth from the SFR alone $z<2$, in agreement with our results.  We refer the reader to Section 5.1 for a discussion of the differences in the observed stellar mass estimates.



\subsection{Evolution in Specific Star Formation Rates}

In the previous two sections we separately examined the cosmically-averaged evolution of the mass and star formation rates of cumulative number density-selected galaxies, providing new information about how galaxies grow, on average, in stellar mass over time.  In the following section, we examine the relationship between star formation rate and stellar mass on the basis of individual galaxies, which provides a separate, valuable metric for understanding galaxy evolution.

Recent studies have shown that the star formation rates of galaxies are inversely correlated with stellar mass at $z<2$ \citep{Zheng07, Damen09}, consistent with models of cosmic downsizing.  Observations at higher redshift have thus far indicated that the star formation rate normalized by stellar mass, or specific star formation rate (sSFR), of galaxies at a constant mass remains approximately constant with time from $z\sim7$ to $z\sim2$ \citep{Stark09, Gonzalez10, Labbe10a, Labbe10b}.  These observations contrast the steady decrease in sSFR from high redshift, which is expected from the majority of theoretical models \citep{Bouche10, Dave10, Dutton10, Weinmann2011}.   Simulations of an evolving metallicity-dependent star formation rate have been evoked to explain the observed sSFR plateau at $z>2$ \citep[e.g.,][]{KD11}.   Improvements in the dust corrections of high-redshift observations have also been shown capable of raising the measured sSFR of galaxies $z>4$, toward a closer agreement with theory \citep{Bouwens11b}.  However, observations of the globally averaged sSFR at $z>2$ still continue to fall significantly short of the predictions from models.



It is worthwhile to determine whether the sSFR plateau at high redshift persists for galaxies selected at constant number density, rather than at constant mass.  We thus present bin measurements of the specific star formation rates for the same four samples of galaxies selected at constant cumulative number density, shown as a function of redshift, in Figure~\ref{fig:ssfrevol}. In agreement with the results of \citet{Noeske07} and \citet{Daddi07}, we find that the globally averaged sSFR declines dramatically with time from $z\sim2$.  Our results also indicate that this declining sSFR extends to $z=6$, at least for the galaxies in the two lowest cumulative number density (and thus, highest mean mass) bins.  Each of the number density-selected samples declines in sSFR from $z\sim4$ to $z\sim2$, and our data suggests that the trend may extend to $z\sim8$.  However, mass constraints become increasingly difficult at such high redshifts, given our available wavelength coverage  Thus, the likely incompleteness in our data at $z>6$ prohibits drawing definite conclusions from the apparent trend at those extreme redshifts. 

Intriguingly, the measurements for the two lowest number density samples indicate approximately one order of magnitude increase in the globally averaged sSFR from $z=2$ to $z=6$, precisely in agreement with the change in the specific gas accretion rate, which is predicted to scale as (1+z)$^{2.5}$ \citep{ND08}.  We have overplotted this predicted specific gas accretion rate, normalized to fit the amplitude of our measurements at $z>2$ for each sample in Figure~\ref{fig:ssfrevol}.    While incompleteness in our high-redshift measurements complicate the comparison for the samples with log $\Phi_{c}\geq-3.5$ Mpc$^{-3}$, the decline in sSFR agrees exceptionally well for the two lowest number density bins.  In all cases, the observed mean sSFR begins to decline more steeply than the specific gas accretion rate model at times later than the peak epoch of star formation ($z\lesssim2$; see Figure ~\ref{fig:sfrevol}).

In addition to the clear evidence for a decreasing sSFR from $z\sim6$ to $z\sim2$, we find a number density-dependent trend: at any given redshift, the sSFR is higher for galaxies at lower number density.  This effect can qualitatively explain the fact that previous results \citep[e.g.,][]{Stark09, Gonzalez10, Labbe10a, Labbe10b, Bouwens11b}, which considered the evolution of the sSFR of galaxies at constant stellar mass, observe a plateau at $z>4$, while we find a constantly declining sSFR from $z\sim6$.  A constant mass selection will effectively probe lower number-density samples with increasing redshift.  As our findings suggest that the sSFR decreases with decreasing number density at all redshifts examined in this work, a plateau in the sSFR evolution would naturally result from a selection at constant mass.  Thus, our findings are not in disagreement with previous results.  Instead, we propose that the method of galaxy selection we employ better illuminates the sSFR evolution of matched galaxy samples over the same redshift range.

We note that our results are also consistent with the previous measurements of specific star formation rates measured for galaxies grouped cumulatively in mass at various redshift intervals in \citet{Damen09}.  Still, due to large differences in the redshift and mass limits of the two samples, only one data point in our respective analyses may be directly compared.   \citet{Damen09} measure an average sSFR of 2 $\times$10$^{-10}$ yr$^{-1}$ for galaxies with log$(M_{*}/M_{\odot})>11$ at $z=1.5$, precisely in agreement with the sSFR we measure for galaxies matched to the same mass and redshift interval.

\begin{figure}[t!]
\centering
\begin{center}
\epsscale{1.25}
\plotone{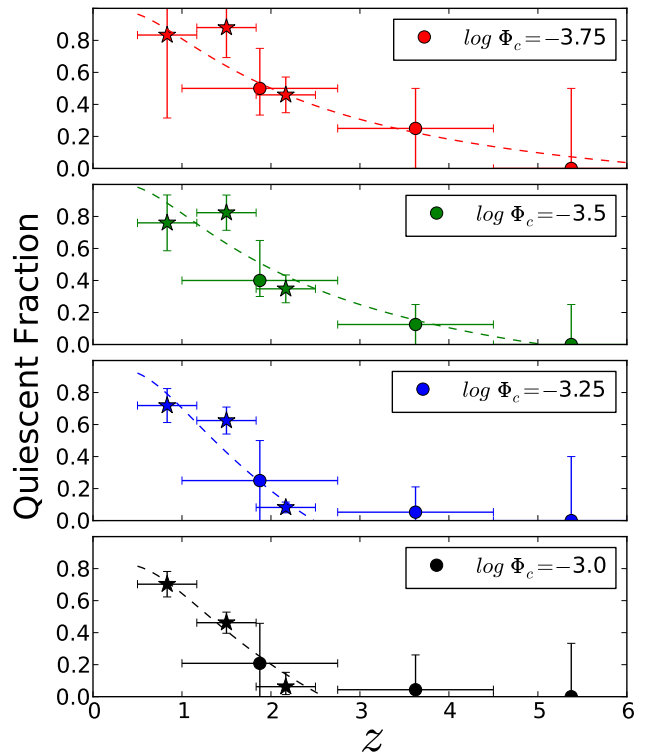}
\caption{The evolution of the fraction of quiescent galaxies, shown for four samples of galaxies with constant cumulative number density.  Here we consider quiescent galaxies to be those with specific star formation rates less than $1/3t_{H}$.  As in previous figures, measurements using the $H_{160}$-selected catalog are plotted as circles;  lower redshift measurements marked with stars are derived using the FIREWORKS catalog of \citet{Wuyts08}. Best-fit exponential curves are overplotted in each panel. \label{fig:qfrac}}
\end{center}
\end{figure}

\section{Evolution the Quiescent Fraction of Galaxies}

Specific star formation rates can additionally provide direct measurements of the star-forming or quiescent nature for individual galaxies.  Thus, a separate averaging of this information can be used to probe an additional aspect of cosmically-averaged galaxy evolution: the changing fraction of quiescent galaxies with time.  

Galaxies in the local Universe exhibit a clear bimodality in their star formation properties, which is most prominently expressed in their color-magnitude distributions \citep[e.g.,][]{Kauffmann03, Baldry04}.  Massive quiescent galaxies have been shown to populate a ``red sequence" in these metrics to $z\sim2$, although this distribution suffers somewhat from contamination by galaxies with obscured star-formation.  Dusty star forming galaxies can be largely separated from the truly quiescent populations by plotting their rest-frame color distributions in U-V vs. V-J \citep{Labbe07, Wuyts07, Williams09}. Using this method, \citet{Whitaker11} revealed the persistence of color bimodality out to $z\sim3$.   \citet{Brammer09} demonstrated that one can also correct the colors of galaxies for the dust reddening using measurements in the infrared, which they used to quantify the fraction of massive quiescent galaxies at $z\sim2.5$.  Despite these advances, the evolution of the fraction of quiescent galaxies remains to be precisely measured at higher redshift, and for galaxies selected at constant number density.

Specific star formation rates provide an alternate method of quantifying the quiescent nature of galaxies.  Under the assumption that galaxies with a sSFR $<\frac{1}{3t_{H}}$ may be considered quiescent, \citet{Damen09} traced the quiescent fraction of galaxies with $M_{*}>10^{11}M_{\odot}$ from $0.5<z<1.75$.  They determined that $19\pm9$\% of galaxies at $z\sim1.8$ could be defined as quiescent.  The number density-selected galaxy samples available in this work provide an ideal sample for confirming this result and extending the quiescent fraction evolution measurements to even higher redshift.  


In Figure~\ref{fig:qfrac}, we present measurements of the quiescent fraction for the same four galaxy samples previous selected at constant cumulative number density.   In keeping with the conventions of \citet{Damen09}, we have employed a requirement of sSFR $<1/3t_{H}$ to identify quiescent galaxies.  Low-redshift measurements calculated using the FIREWORKS catalog are overplotted as stars.  We assume Poisson statistics when estimating the error on the FIREWORKS measurements.  Errors on these fractional measurements for the HUDF catalog are non-Gaussian due to the small sample size.  These errors have therefore been determined by sampling the 1$\sigma$ confidence intervals of the individual measurements but may still be somewhat underestimated due to the further complications of cosmic variance. 

Despite the poorly constrained statistics in the HUDF sample, we find a continually declining quiescent fraction with increasing redshift, which becomes consistent with zero for all sub-samples beyond $z\sim4$.  We also find that the quiescent fraction increases with decreasing number density, at any given redshift.  These measurements are largely consistent with \citet{Damen09}, although again, due to the small overlap in the limiting redshift and mass ranges, we can only compare our data for log $\Phi_{c}=-3.5$ Mpc$^{-3}$ at $z=1.5$.  Here, \citet{Damen09} finds $\sim40$\% of galaxies to be quiescent, whereas we find approximately 60\%.  In both cases, the 1$\sigma$ errors on the measurements are large enough to consider these findings to be consistent.




\section{Summary}

Using recent deep imaging from the WFC3/IR camera on {\it HST} we have constructed a $H_{160}$-selected catalog of galaxies in the Hubble Ultra-Deep Field, with a limiting magnitude of $H_{160}=27.8$.  By incorporating archival deep imaging from the UV, optical, and infrared, we have produced reliable SEDs with complete rest-frame UV and optical coverage for $\sim1500$ galaxies, from which we estimate the redshift, and extract stellar masses and star formation rate estimates from stellar population synthesis modeling.  Using these estimates we measure the cumulative mass function from $z=6$ to $z=1$ and produce four samples of galaxies selected with constant cumulative number densities ranging from log $\Phi_{c}=-3.0$ Mpc$^{-3}$ to log $\Phi_{c}=-3.75$ Mpc$^{-3}$.  This sample selection enables an analysis of the properties of galaxies and their progenitors across the wide range in redshift, while avoiding the significant biases produced by other means of selection (e.g., color, mass, luminosity).  

Our examination of the evolution of the galaxies at constant cumulative number density produced the following key results:

\begin{enumerate}

\item We observe an exponential growth in stellar mass equivalent to $\sim1.5$ magnitudes from $z=6$ to $z=2$ for galaxies selected with constant cumulative number densities of  log $\Phi_{c}=-3.0$, $-$3.25, $-$3.5, and $-$3.75 Mpc$^{-3}$.  

\item The mean stellar mass evolution we measure for galaxies with log $\Phi_{c}=-3.75$ Mpc$^{-3}$ at $z>2$ is consistent with the results from \citet{Papovich11} obtained using the mass-selected sample of \citet{Marchesini09}.  While still agreeing within the stated 1$\sigma$ errors, our mean stellar mass measurements are consistently a factor of $\sim$2 higher than those inferred strictly from the integrated luminosity function by \citep{Papovich11}.  This difference in results suggests that the H$_{160}$ selection and supplementary IRAC coverage equip our sample to achieve higher completeness with respect to galaxies at lower star formation rates.

\item We observe a gradual increase in the star formation rates of galaxies at constant number density from $z\sim6$ to $z\sim3$.  The star formation rate evolution we measure at $z>2$ is consistent with other measurements \citep{Papovich11, Smit2012} and agrees well over our full range of selected number densities with the model predictions of \citet{KD11}.  We find that each of the number density-selected samples reaches a peak SFR of $\sim100$ $[M_{\odot} yr^{-1}]$ at $2\leq z\leq3$, and that the epoch of peak SFR shifts to later times for galaxies with increasing number densities, in agreement with the expectations from cosmic downsizing.

\item By comparing the observed mean mass evolution with the observed mean star formation rate evolution of galaxies selected for a range of cumulative number densities, we find that the mean star formation rates of galaxies with log $\Phi_{c}\geq-3.25$ Mpc$^{-3}$ can fully account for the mean growth in stellar mass from $0.5<z<5$.  However, at lower constant cumulative number density (higher mean mass), the observed star formation rates are unable to produce the observed buildup of stellar mass, falling short by a factor of $\sim3$ at $z<2$.  While mergers almost certainly play a role, particularly at late times, in building up the difference between the observed stellar mass growth and what is expected from the measured star formation rates, obscured star formation may also make a significant contribution.

\item An analysis of the mean evolution of the specific star formation rates for galaxies with  log $\Phi_{c}\leq-3.5$ provides strong evidence of a steadily declining sSFR from $z=6$ to $z=2$ equivalent to one order of magnitude, in agreement with the predicted evolution of the specific gas accretion rate during this time \citet{ND08}.  We additionally find that, at any given redshift, the cosmically averaged sSFR is higher for galaxies with higher number densities (lower mean stellar mass).  We suggest that the ``sSFR plateau" found in previous analyses, which considered the evolution of galaxies with constant mass, at $z>4$ can be qualitatively explained by the combination of these findings.

\item Using the estimated sSFRs of individual galaxies, we measure the increase in the global fraction of quiescent galaxies from $z\sim6$ as a function of number density.  These results are poorly constrained at $z>3$ due to our small number statistics, but they broadly agree with previous measurements.  We additionally find no evidence in our small sample for quiescent galaxies at $z>4$.
\end{enumerate}

\section{Future Work}

While the new and archival multi-wavelength observations of the Hubble Ultra-Deep Field have enabled unparalleled insights into the high-redshift Universe,  the angular size of the HUDF is also exceptionally, and somewhat prohibitively, small.   Thus, the effects of cosmic variance substantially contribute to the errors in the observations presented in this work.  Future extensions of this analysis that include data from the HUDF05 flanking fields \citep{Oesch07}, for which similarly deep ACS and WFC3/IR observations are now available,  will help to reduce the statistical errors from both poisson variation and cosmic variance.  Repeating this analysis with the most recent HUDF12 imaging \citep{Koekemoer12, Ellis13} will also help to push the data within the field to even fainter limits.

As we have shown, the inclusion of deep IRAC photometry is also critical for constraining the stellar mass estimates of detected galaxies.  New ultra-deep Spitzer/IRAC observations of these flanking fields from the 262 hour IUDF program \citep[PI: Labb{\'e},][]{Labbe12} will reduce the effects of cosmic variance by a factor of 2 and reduce the uncertainties in stellar masses in the high-redshift sample by a factor of 2-4.  Exciting results have already come from analyses of the larger Early Release Science (ERS)  \citep{Windhorst2011} and CANDELS Deep and Wide fields \citep{Grogin11, K11}. Larger samples, made available by these and other deep fields will provide better constraints on the mass and star formation rate evolution explored in this work, as well as enabling a higher resolution and more expansive exploration of how the evolutionary tracks vary with number density.

\acknowledgments

We thank the anonymous referee for many insightful comments and suggestions, which have been incorporated into this work.  We are additionally grateful to Adam Muzzin, Danilo Marchesini, Ryan Quadri, David Wake, Kate Whitaker, and Simone Winemann for helpful and enlightening discussions.  We acknowledge support from HST-GO10937, HST-GO11563, HST-GO11144, HST-GO12177, and from ERC grant HIGHZ no. 227749.  BL acknowledges support from the National Science Foundation Astronomy and Astrophysics Fellowship grant AST-1202963.  PO acknowledges support by NASA through Hubble Fellowship grant HF-51278.01-A, awarded by the Space Telescope Science Institute, which is operated by the Association of Universities for Research in Astronomy, Inc., for NASA, under contract NAS 5-26555.



{\it Facilities:} \facility{HST (STIS)}, \facility{SSO (SSC)}.



\appendix
\section{A. Description of the Photometric Catalog}

The H$_{160}$-selected catalog described in Section 3 and used throughout our analysis is available in the electronic version of this publication.  We describe the format of this catalog in Table 3.  The catalog fluxes can be converted to total magnitudes as follows:
\begin{equation}  
m = 25 - 2.5\log(F_{n})
\end{equation}
where $F_{n}$ is the total flux density in each filter for the object with a magnitude m, normalized to a zero point of 25 in the AB system, corresponding to a flux density of $3.631\times10^{-30}$ erg s$^{-1}$ Hz$^{-1}$ cm$^{-2}$.

Source extraction output parameters from SExtractor determined using the H$_{160}$ detection image are provided in fields 28 and 30.  Field 29 gives the fraction of maximum exposure in the H$_{160}$ image at the center of each object, as determined using the H$_{160}$ weight map.  Fields 32 and 33 respectively provide spectroscopic and photometric redshifts from the literature, when available.  Previous spectroscopic redshifts listed in the catalog are drawn from the GRAPES \citep{Pirzkal04} and PEARS \citep{Straughn08} catalogs.  The previous photometric redshifts refer to those included in the FIREWORKS catalog \citep{Wuyts08}. The photometric redshifts and errors determined with EAZY are included in Fields 34-36.  

After visual inspection 20 objects have been excluded from the analysis in this work, due to their catastrophic proximity to bright objects, edges, or other artifacts.  These objects are marked with a value of 0 in the `useflag' field

\begin{figure}[]
\centering
\begin{center}
\epsscale{1.1}
\plotone{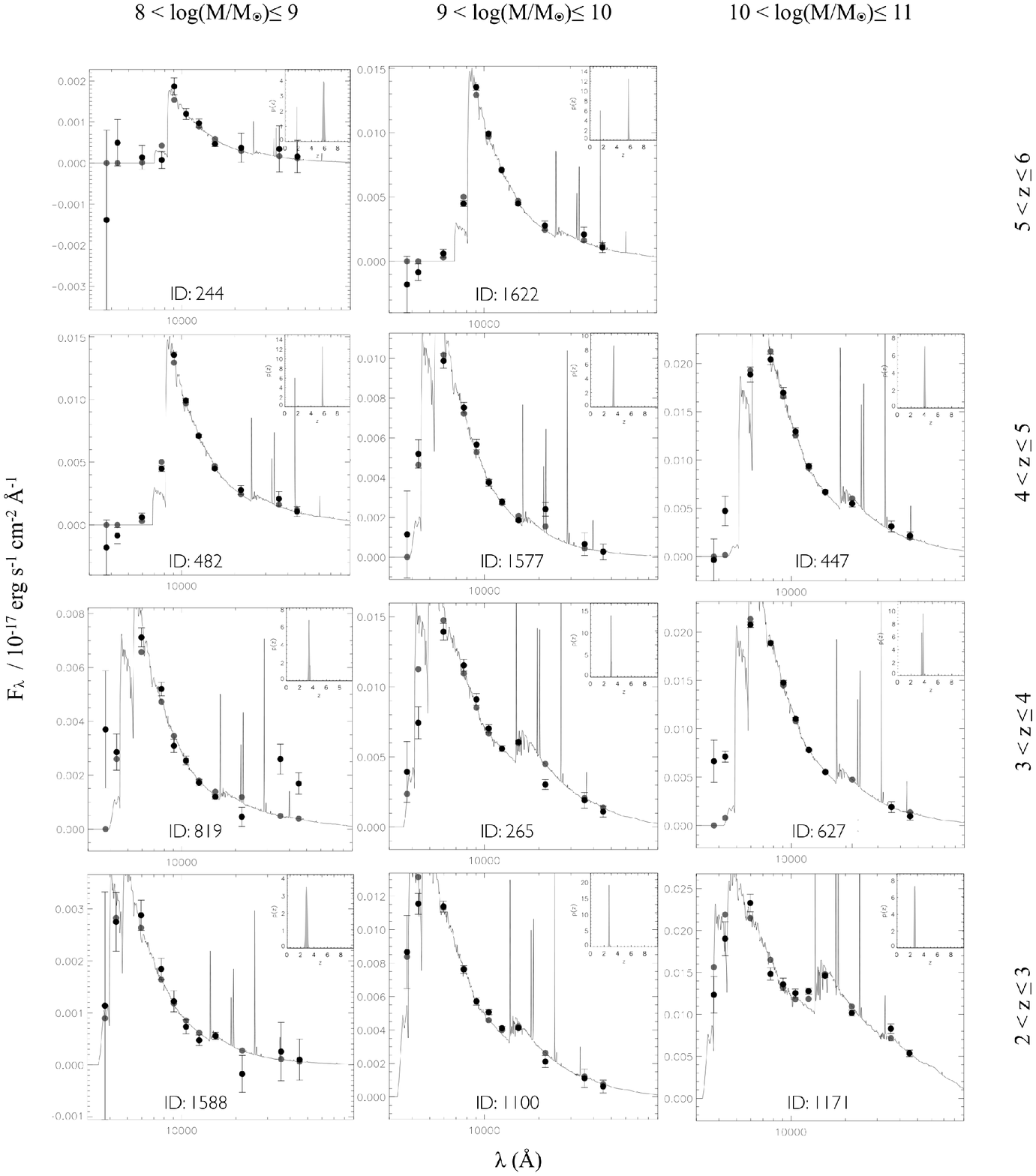}
\caption{Example SEDs with a photometric redshift precision of at least 20\% in the range $2<z\leq6$, have been drawn at random from the catalog to fill a grid in redshift and stellar mass. Inset in the upper-right corner of each panel is a plot of the redshift probability distribution, P(z), determined by EAZY.  Flux density measurements from the catalog are presented with black dots.  The best-fit template spectrum from EAZY is over plotted in gray, with gray dots marking the best-fit template flux at the central wavelength of each filter. \label{fig:exampleseds}}
\end{center}
\end{figure}

Example photometric SEDs and best fit model spectra with a photometric redshift precision of at least 20\% in the range $2<z\leq6$ have been randomly selected from the catalog and presented in Figure \ref{fig:exampleseds} in order of increasing mass (left to right) and redshift (bottom to top).

\begin{table*}
\centering
\begin{minipage}{0.75\textwidth}
\begin{center}
\caption{Photometric Catalog Format}
\begin{tabular}{@{}rrr@{}}
\hline
Column & Header & Description \\
\hline
\hline
  1& ID & Unique object identifier \\
  2, 3 & F103, E103 & U$_{38}$ Flux, Error \\
  4, 5 & F1, E1 & F435W Flux, Error \\
  6, 7 & F4, E4 & F606W Flux, Error \\
  8, 9 & F6, E6 & F714W Flux, Error \\
  10, 11 & F7, E7 & F850LP Flux, Error \\
  12, 13 & F202, E202 & F125W Flux, Error \\
  14, 15 & F203, E203 & F140W Flux, Error \\
  16, 17 & F205, E205 & F160W Flux, Error \\
  18, 19 & F37, E37 & K$_{s}$-Band Flux, Error \\
  20, 21 & F18, E18 & 3.6$\mu$m Flux, Error\\
  22, 23 & F19, E19 & 4.5$\mu$m Flux, Error \\
  24, 25 & ra, dec & Right Ascention, Declination \\
  26, 27 & x, y & X , Y image coordinates \\
  28 & blendflag & SExtractor blend flag \\
  29 & weight$_H$ & Weight of the F160W exposure \\
  30 & aprad$_{Hauto}$ & SExtractor F160W auto aperture radius \\
  31 & apcor &  Factor of aperture correction to total \\
  32 & z$_{spec}$ & Previous spectroscopic redshift \\
  33 & z$_{phbest}$ & Previous best photometric redshift \\
  34 & z$_{peak}$ & Peak of the EAZY P(z)\\
  35, 36 & z$_{l68}$, z$_{u68}$ & Lower, upper 1$\sigma$ EAZY limits on z$_{peak}$ \\
  37 & useflag &  Flag marking catastrophic blends or artifacts\\
\hline
\end{tabular}
\end{center}
\end{minipage}
\end{table*}

\section{B. Details of the IRAC Photometry}



\begin{figure}[t!]
\centering
\begin{center}
\epsscale{0.7}
\plotone{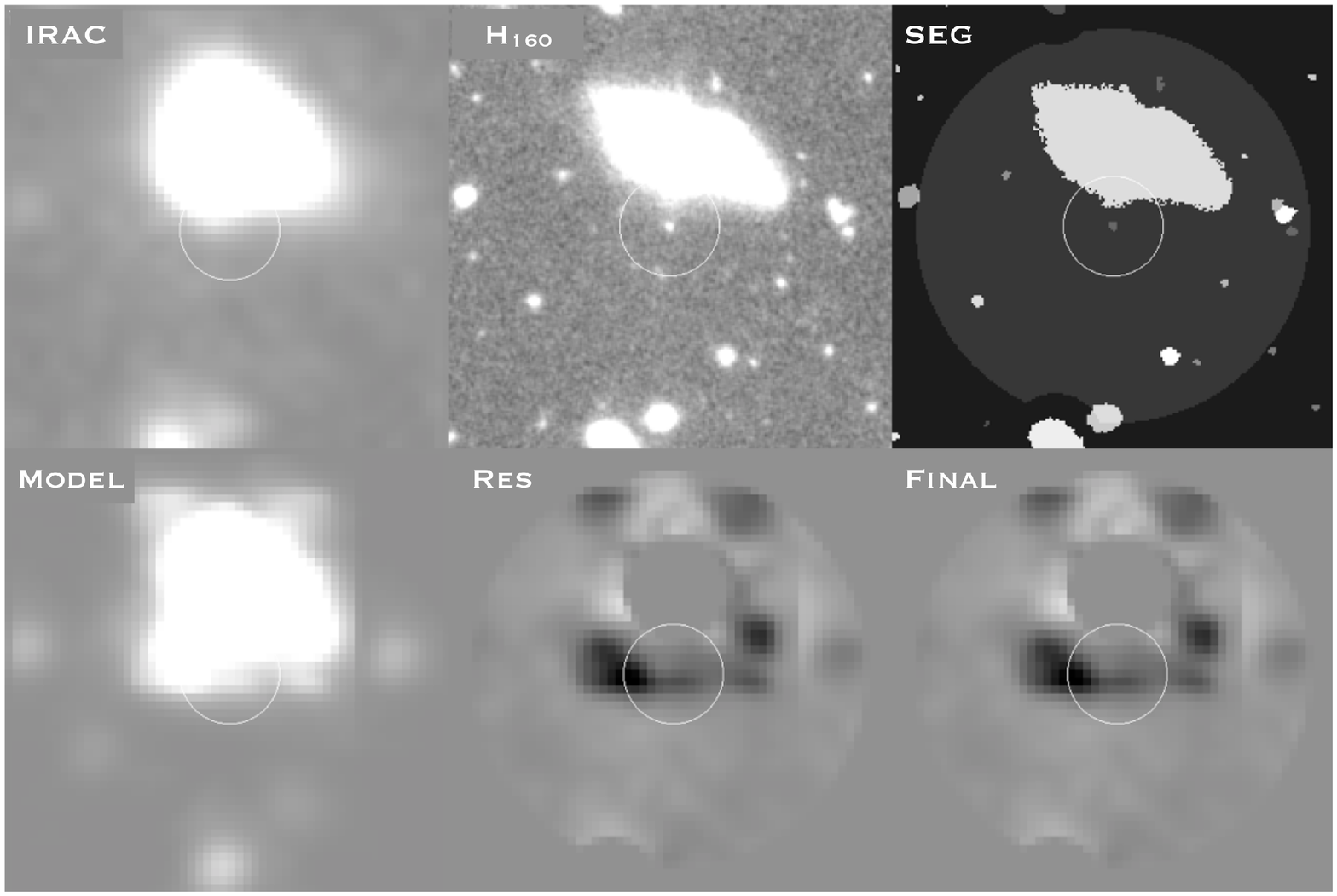}
\caption{An example of sub-optimal subtraction of a bright 3.6$\mu$m source in close proximity to a faint galaxy detected in F140W (encircled).  The original IRAC 3.6$\mu m$ cutout image for catalog object ID 877, which has been excluded from our scientific analysis, is shown top left.  The top center panel gives the matching $H_{160}$ detection image, and the segmentation map of the region from \textsc{SExtractor} is shown at top right.  The bottom panel shows, from left to right, the modeled IRAC flux for all objects in the region, the residual image with all modeled fluxes removed, and the flux for the central object alone. \label{fig:badsubphot}}
\end{center}
\end{figure}

The PSF matching technique described in Section 3.2 has enabled us to directly combine the aperture photometry from images with dramatically different PSFs.  This task is essential for the inclusion of IRAC photometry, which has exceptionally broad-winged PSFs, complicating the subtraction of closely spaced sources detected in the deep F160W image.  Visual inspection of the residual source-subtracted images for each galaxy suggests that the procedure we employ to fit and subtract the light profiles of neighboring sources works very well for majority of objects in the field (see, for example, Figure ~\ref{fig:subphot}).  However, in the cases of the brightest IRAC sources, subtle imperfections in the PSF determination can produce suboptimal fitting for nearby sources.  We present an example of such a case in Figure ~\ref{fig:badsubphot}.

To investigate the nature of this effect, we superimposed 4000 artificial point sources to the F160W detection image at random positions.  We then ran the PSF matching algorithm on the original IRAC 3.6$\mu$m image and examined the output aperture-corrected total magnitudes as a function of distance and flux from nearby neighbors.  If the algorithm worked perfectly, the empty apertures would return values equivalent to the limiting magnitude of the background ($\sim27$).  Instead, we find that the magnitudes measured within these apertures contain significant excess absolute flux when placed in close proximity to large, bright sources.

\begin{figure}
\centering
\begin{center}
\epsscale{1.1}
\plotone{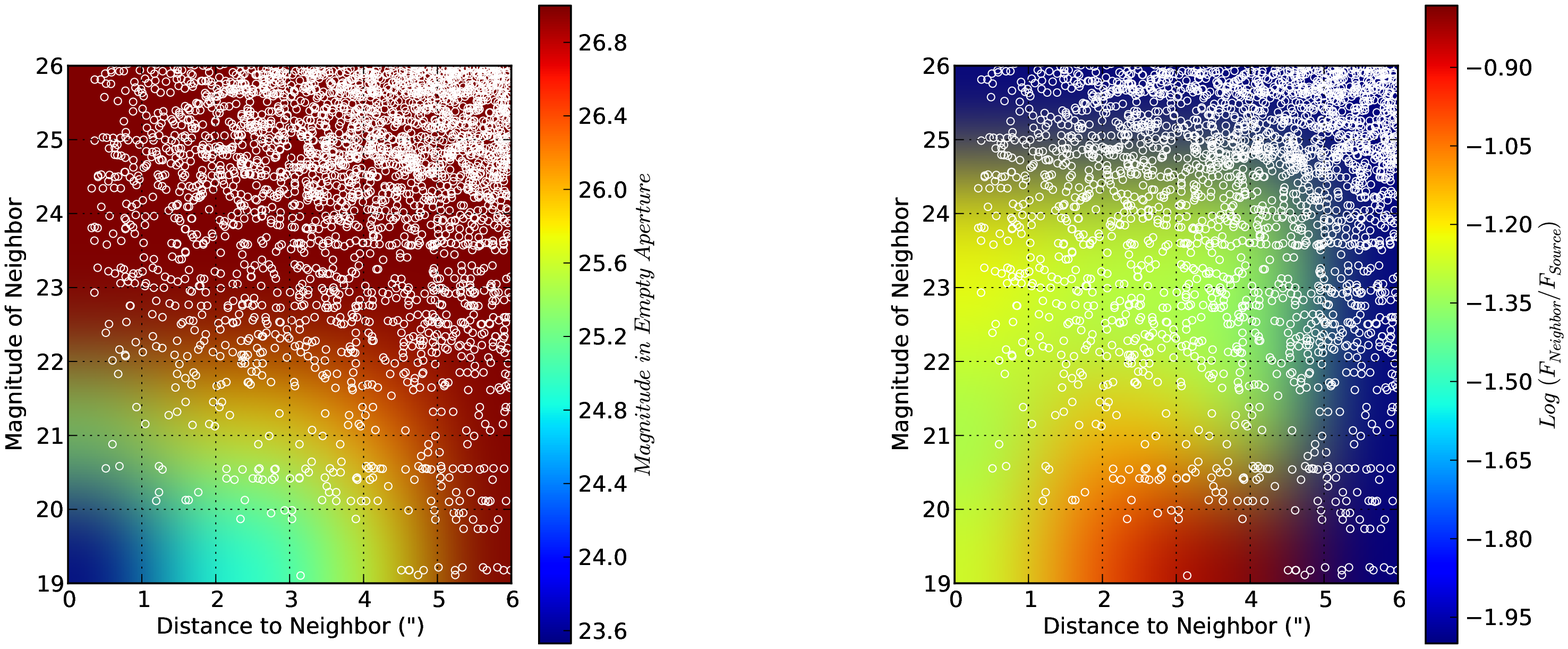}
\caption{{\it Left:} The separation of galaxies with $\geq2\sigma$ IRAC detections retained in the HUDF catalog, as a function of the total 3.6$\mu$m magnitude of one neighbor (white circles). The underlying color contours present the residual magnitude measured in randomly placed empty apertures as a function of neighboring source magnitude and distance after PSF-matching the F160W detection image with the IRAC 3.6$\mu$m imaging.   {\it Right:} The same data points, this time with the underlying color representing the mean fractional flux contamination of individual galaxies by neighbors.  In the worst cases, the contamination reaches a level of 15\%.  Sources with estimated fractional flux contamination higher than this level have been excluded from the analysis.  \label{fig:ivotest}}
\end{center}
\end{figure}

In Figure ~\ref{fig:ivotest} we present the separation of neighboring galaxies in our catalog as a function of total magnitude of one neighbor.  Underlying this distribution is the interpolated magnitude detected in the empty apertures randomly placed around the objects in the catalog, as a function of both magnitude and distance of the neighboring galaxy.   We find that significant contamination persists out to 5" around galaxies with IRAC 3.6$\mu$m magnitudes brighter than 22.  A total of 32 galaxies in the HUDF have magnitudes brighter than this limit, and their neighboring galaxies constitute $\lesssim2$\% of the total number of pairs with similar separations.  

As the significance of the contamination from neighbors will depend on the intrinsic flux of the primary source, we replot the data with color scaling to represent the estimated fractional flux contamination in Figure ~\ref{fig:ivotest}.  Neighboring galaxies with IRAC 3.6$\mu$m magnitudes fainter than 22 produce less than 10\% flux contamination, regardless of angular separation.  Cases of catastrophic blending have been flagged and removed from the analysis (see the description of the `useflag' field of the catalog in Appendix A).

\clearpage

\end{document}